\documentclass{article}

\usepackage{graphicx}
\usepackage{natbib}
\usepackage{multirow}
\usepackage{booktabs}
\usepackage{amsmath}
\usepackage{amssymb}
\usepackage{amsfonts}
\usepackage[dvipsnames]{xcolor}
\usepackage{moreverb}
\usepackage{multirow}
\usepackage[geometry]{ifsym}
\usepackage{MnSymbol}
\usepackage{pifont}
\usepackage{lpic}
\usepackage{rotating}
\usepackage{xfrac}
\usepackage{fullpage}

\newsavebox{\astrutbox}
\sbox{\astrutbox}{\rule[-5pt]{0pt}{20pt}}

\title{The non-equilibrium region of grid-generated decaying turbulence}
\author{P. C. Valente and J. C. Vassilicos \\
Department of Aeronautics, Imperial College London,
London SW7 2AZ, United Kingdom}%

\date{}

\begin{document}

\maketitle

\begin{abstract}
The previously reported non-equilibrium dissipation law is
investigated in turbulent flows generated by various regular and
fractal square grids. The flows are documented in terms of various
turbulent profiles which reveal their differences. In spite of
significant inhomogeneity and anisotropy differences, the new
non-equilibrium dissipation law is observed in all these
flows. Various transverse and longitudinal integral scales are
measured and used to define the dissipation coefficient
$C_{\varepsilon}$. It is found that the new non-equilibrium
dissipation law is not an artefact of a particular choice of the
integral scale and that the usual equilibrium dissipation law can
actually coexist with the non-equilibrium law in different regions of
the same flow.

\vspace{5mm}
\end{abstract}

\section{Introduction}
The classical empirical scaling, $C_{\varepsilon} \equiv \varepsilon
\ell/u'^3 \approx \mathrm{constant}$ (where $\varepsilon$, $\ell$ and
$u'$ are, respectively, the turbulent kinetic energy dissipation per
unit mass, an integral length-scale and the root-mean-square of the
fluctuating velocity field) was first suggested by \cite{Taylor1935}
and is considered to be ``one of the cornerstone assumptions of
turbulence theory'' \cite[][pp. 20-21]{TennekesLumley:book}.
The importance of $C_{\varepsilon} \approx \mathrm{constant}$ can
hardly be overstated since it is often used in the modelling of
turbulence and as a building block in the understanding of turbulence
physics \cite[see
  e.g.][]{Townsend:book,TennekesLumley:book,Frisch:book}.  This
phenomenological relation has received support conceptually \cite[see
  e.g.][]{Lumley92,Frisch:book}, experimentally
\cite[]{Sreeni84,Sreeni95,Pearson,Burattini2005} and numerically
\cite[]{Sreeni98,Burattini2005}, for both homogeneous and
inhomogeneous flows, with or without mean shear, whether stationary or
non-stationary.

However, the general validity of this empirical law has been strongly
challenged over the past six years. Wind tunnel and water flume
experiments on turbulence generated by fractal square grids (FSGs)
using both hot wire anemometry (HWA) and particle image velocimetry
(PIV) \cite[]{SV2007,MV2010,VV2011,VV2012,gomesfernandesetal12,discettietal11,Nagata2012} have provided a wealth of data for which
$C_{\varepsilon} \approx \mathrm{constant}$ does not hold and is
replaced by $C_{\varepsilon} = Re_{M}^{m}/Re_{\ell}^{n}$ where
$m\approx n \approx 1$ seem to be good approximations supported by the
data even at fairly high Reynolds numbers. ($Re_{\ell} = u' \ell/\nu$
and $Re_M = U_{\infty} M/\nu$ where $U_{\infty}$ is the inlet
velocity, $M$ is an inlet mesh size and $\nu$ is the fluid's kinematic
viscosity.)

\cite{VV2012} showed that this non-classical
dissipation behaviour is also present in a flow region of decaying
turbulence generated by regular square-mesh grids (RGs) closer to the
grid than the far region where the classical $C_{\varepsilon} \approx
\mathrm{constant}$ behaviour is recovered. The classical and
non-classical behaviours of $\varepsilon$ were associated with
equilibrium and non-equilibrium turbulence, respectively, 
since, as shown in textbooks such as Pope (2000, pp. 182 -- 188), the classical
behaviour $C_{\varepsilon}=\mathrm{constant}$ is a consequence of the
Richardson-Kolmogorov equilibrium cascade.  This
observation is significant because it shows (i) that the
non-equilibrium behaviour occurs naturally, and possibly generally, in
nature without being particular to the FSGs and (ii) that the non-equilibrium
behaviour can lead to the equilibrium behaviour along a streamwise
evolution.
However, the -5/3 power-law of the energy spectrum is already
present in its clearest and over the longest range in the non-equilibrium
region.

In this study we investigate different forms of the $C_{\varepsilon}$ law in terms of
different definitions of the integral length-scale which is usually
taken to be the longitudinal integral length-scale but does not need
to be. We try different longitudinal/lateral integral length-scales
and also investigate dependencies on isotropy which can impact on the
use of $3/ 2\, u'^{2}$ as a surrogate for kinetic energy. We also
examine the sensitivity of our results on the way that the dissipation
rate is obtained. 

It is also important to further document the non-equilibrium
region of the flow as a platform for such an investigation. 
We therefore start by doing this for two RGs and two FSGs, and
we obtain and compare various turbulent flow profiles in the vertical
mid-plane using two-component hot wire anemometry (\S
\ref{sec:topology}). These data allow a quantitative assessment of the
turbulent transport and production (\S \ref{sec:homo}) and help us
assess the impact on the flow of the geometrical differences between
the grids. The data also allow a qualitative assessment of the
confinement due to the wind tunnel's bounding walls, \S\S
\ref{sec:confinement}, \ref{sec:homo}. The turbulence generated by one
of the grids, RG115 (figure \ref{fig:grids}d) is then further
investigated using two-point/two-component anemometry measurements to
estimate various integral length-scales of the flow. These
measurements also allow a quantitative assessment of large-scale
anisotropy (\S \ref{sec:Lu}) and are finally used to investigate, for
the first time, the behaviour of the normalised energy dissipation
using various integral length-scales. Both centreline and
off-centreline assessments, in particular behind a bar of the grid,
are performed (\S \ref{sec:Llambda}).

\section{Experimental setup}\label{sec:apparatus}
The experiments are performed in the 3'x3' closed circuit and the 18''x18'' blow-down wind tunnels at the Department of Aeronautics, Imperial College London. The 3'x3'  wind tunnel has a working section of 0.91 m x 0.91 m x 4.8 m, a contraction ratio of 9:1 and the free stream turbulence intensity is about 0.05 \%. The 18''x18'' wind tunnel has a working section of 0.46 m x 0.46 m x 3.5 m,  a contraction ratio of 8:1 and the free stream turbulence intensity is about 0.1 \%.
(Note that, the length of the test section of the 18''x18'' is the same as that used in \citealt{VV2012}, which is about $1$m shorter than that used in \citealt{VV2011}.)
 The inlet velocity $U_{\infty}$ in both tunnels is set and stabilised with a PID feedback controller which takes as an input the static pressure difference across the contraction and the flow temperature (both measured using a Furness Controls micromanometer FCO510) and actuates on the input of the wind tunnel's motor drive.

\begin{figure}
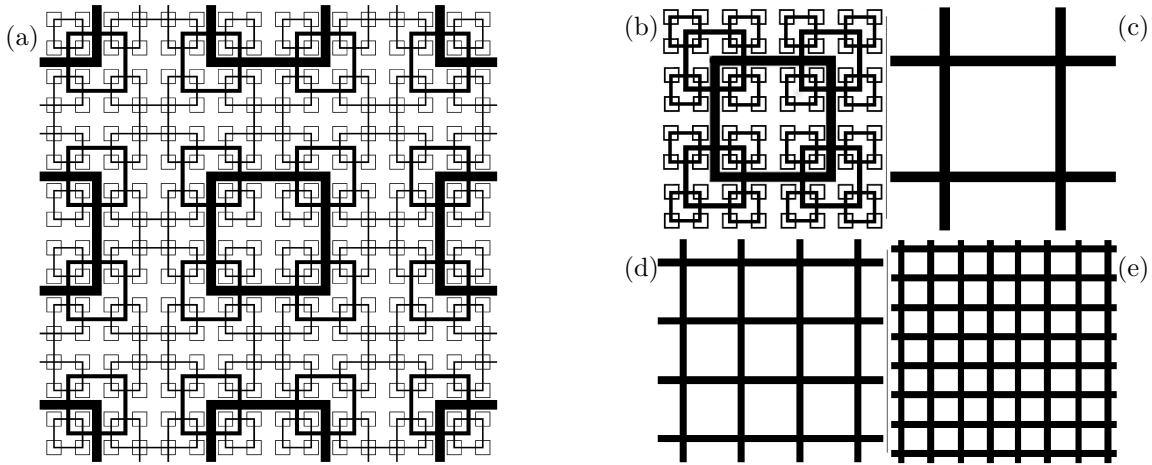

\begin{minipage}[c]{0.5\linewidth}
   \centering
   \begin{lpic}{Figures/FSG3x3(62mm)}
   \lbl{-5,160;(a)}
   \end{lpic}
\end{minipage}%
\begin{minipage}[c]{0.5\linewidth}
   \centering 
   \begin{lpic}{Figures/Grids18in(62mm)}
   \lbl{-6,175;(b)}
   \lbl{192,175;(c)}
   \lbl{-6,78;(d)}
   \lbl{192,78;(e)}
   \end{lpic}
\end{minipage}
\caption{Turbulence generating grids, (a) FSG 3'x3' (b) FSG 18'' (c) RG230 (d) RG115 and (e) RG60. 
The figures are to scale.}
\label{fig:grids}
\end{figure}

Data are recorded at the lee of five grids (figure \ref{fig:grids}). 
The geometrical details of the grids are summarized in table \ref{table:grids} and the overview of the data acquired  for each grid are presented in table \ref{table:summary}.
 All our grids were used in the 18''x18'' wind tunnel with the sole exception of the FSG3'x3' grid which was used in the 3'x3' wind  tunnel. 
The coordinate system that we use has the streamwise centreline as $x$-axis and two axes orthogonal to each other and to the centreline but aligned with the bars on the grids as $y$- and $z$-axes. 
The origin of this coordinate system is on the centreline at the test section entry where the grid is placed. 
The streamwise and two spanwise velocity components are $U_{1} + u_{1} = U+u$, $U_{2} + u_{2} = V+v$ (aligned with the $y$-axis) and $U_{3} + u_{3}= W+w$ (aligned with the $z$-axis) where $U_{1}= U$, $U_{2}= V$ and $U_{3}=
W$ are mean flow components and $u_{1}=u$, $u_{2} = v$ and $u_{3}= w$ are turbulence fluctuating velocity components.

\begin{table}
\centering
\begin{tabular*}{0.9\textwidth}{@{\extracolsep{\fill}}lccccccc}
Grid &  & $M$ & $t_0$ & $d$ & $\sigma$ & $x_*$ & $x_{\mathrm{peak}}/x_*$ \\
        &  & (mm) & (mm) & (mm) & (\%) & (m)&  \\
\midrule
FSG3'x3'     & mono-planar &  228.7 & 19.2 & 5   & 25 & 2.72 & 0.43\\ 
FSG18''x18'' & mono-planar & 237.7 & 19.2 & 5   & 25 & 2.94 & 0.43\\ 
RG230 & mono-planar & 230    & 20    & 6   & 17 & 2.65 & 0.63\\
RG115 & mono-planar & 115    & 10    & 3.2   & 17 & 1.32& 0.63 \\
RG60   & bi-planar       & 60      & 10    & 10 & 32 & 0.36 & $\simeq 0.4$\\
\end{tabular*}
\caption{Geometric details of turbulence-generating grids. 
For the RGs, $M$ is the distance between parallel bars within a mesh, i.e. the mesh size, and $t_0$ is the lateral thickness of these bars. 
For the FSGs, which are made of many different square bar arrangements, i.e. meshes, of different sizes, $M$ refers to the largest mesh size and $t_0$ to the lateral thickness of the largest bars. 
The longitudinal thickness of the bars is denoted as $d$, and $\sigma$ is the grid blockage ratio. 
The value of $x_{\mathrm{peak}}$ for RG60 (see \S \ref{sec:wake}) is taken from measurements of a very similar grid 
($x_{\mathrm{peak}}$ and $x_*$ are defined in  \S \ref{sec:wake}).
The low-blockage space-filling FSGs, have four fractal iterations and a thickness ratio $t_r$, i.e. the ratio between the lateral thickness of the biggest to the lateral thickness of the smallest bars, equal to $17$. For further details see \cite{VV2011}.}
\label{table:grids}
\end{table}

\begin{table}
\centering
\begin{tabular*}{\textwidth}{@{\extracolsep{\fill}}rcclcccl}
Grid & $U_{\infty}$ & $Re_M$ & Probe & $x/x_{\mathrm{peak}}$ & $y/M$ & Dataset & \# \\
    & (m/s)  & ($\times 10^3$)& & & &  \\
\midrule
 FSG3'x3' & 15 & 229 & SW$2.5{\mu}$m & [0.7 -- 3.2] & [0; 0.5] & New & 1\\
FSG3'x3' & 15 & 229 & XW$5{\mu}$m & [1.5; 2.0; 2.5; 3.0; 3.5] & [-0.5 -- 0.5] & New & 2\\
FSG18''x18''  & 15 & 238 & XW$5{\mu}$m & [1.4; 1.8; 2.3; 2.8; 3.2] & [-0.5 -- 0.5] & VV11 &3  \\
FSG18''x18''  & 15 & 238 & SW$2.5{\mu}$m & [1.3 -- 3.4] & [0] & VV11 & 4 \\
RG230 & 15 & 230 & XW$2.5{\mu}$m & [1.3; 1.8] & [0 -- 0.5] & New &5\\
RG115 & 15 & 115 & XW$2.5{\mu}$m & [1.4; 1.8; 2.8; 3.7] & [-0.5 -- 0.5] & New&6\\
RG115 & 20 & 153 & SW$1{\mu}$m & [0.6 -- 3.8] & [0] & VV12 &7 \\
RG115 & 20 & 153 & SW$1{\mu}$m & [0.6 -- 3.8] & [0.5] & New &8\\
RG115 & 10 & 77   & XW$2.5{\mu}$m($\times 2$)& [1.1; 1.5; 2.0; 2.6; 3.1; 3.7] & [-1.1 -- 1.1] & New&9\\
RG115 & 10 & 77   & XW$2.5{\mu}$m($\times 2$)& [1.1; 1.5; 2.0; 2.6; 3.1; 3.7] & [-1.6 -- 0.6] & New&10\\
RG60 & 10 & 40  & XW$2.5{\mu}$m($\times 2$) & 21 & [-2.1 -- 2.1] & New &11\\
RG60 & 10 & 40  & SW$1{\mu}$m & [1.8 -- 22] & [0] &  VV12 &12 \\
RG230 & 10 & 153  & SW$2.5{\mu}$m & [1.0 -- 1.9] & [0] &  VV12 &13

\vspace{2mm} \\
RG115 & 10 & 77 & XW$2.5{\mu}$m($\times 2$)& [1.5; 2.0; 2.6; 3.1; 3.7] & [-0.6 -- 0.6] & New  & 14\\
RG115 & 10 & 77 & XW$2.5{\mu}$m($\times 2$)& [1.5; 2.6] & [-0.65 -- 0.55] & New & 15\\
RG60 & 10 & 40   & XW$2.5{\mu}$m($\times 2$) & [8.5; 11.5; 15.6; 17.6; 20.7] & [-1.2 -- 1.2] & New  & 16\\
\end{tabular*}
\caption{Overview of the experimental data  \cite[note that VV11 and VV12 refer to][]{VV2011,VV2012}.}
\label{table:summary}
\end{table}

\subsection{Single-point thermal anemometry measurements}\label{sec:singlepoint}
For the single point measurements, i.e. using only one sensor recording one or two velocity components, we use the experimental hardware, setup and instrumentation performance tests described in \cite{VV2011}. 
Briefly, a Dantec Streamline constant temperature anemometer (CTA) is used to drive one- and two-component hot-wires (SW and XW, respectively). 
The majority of the single-point data presented here were recorded with an in-house etched Pl-($10\%$)Rh XW, denoted as XW$2.5{\mu}$m (the sensing length of the wires is $l_w \approx 0.5$mm and their diameter $d_w \approx 2.5\mu$m; the wires are separated by $\Delta \approx 0.5$mm).
 The exceptions are, (i) the two-component data acquired for the FSGs (both the new data and the previous data presented in \citealt{VV2011}, i.e. datasets 2 and 3 respectively) which are recorded with a standard Dantec probe 55P51, denoted as XW$5{\mu}$m ($l_w \approx 1$mm, $d_w \approx 5\mu$m  and the wires are $\Delta \approx 1$mm apart), (ii) the one-component longitudinal profiles at the centreline  and behind the bar at $y = 0.5 M$ of FSG3'x3' (dataset 1) and at the centreline of FSG18''x18'' and RG230 (dataset 4 and 13) which are acquired with an in-house etched Pl-($10\%$)Rh SW, SW$2.5{\mu}$m ($l_w \approx 0.5$mm, $d_w \approx 2.5\mu$m) and (iii) the one-component longitudinal profiles at the centreline of RG115 and RG60 (datasets 7 and 12 from \citealt{VV2012}, respectively) and behind the bar at $y = 0.5 M$ of RG115 (dataset 8) which are acquired with another in-house etched Pl-($10\%$)Rh SW, SW$1{\mu}$m ($l_w \approx 0.2$mm, $d_w \approx 1\mu$m).
The SWs and XWs are mounted, respectively, on a 2- and 3-axes automated traverse mechanism, which sets the downstream and vertical position of the probes and for the XWs also controls the pitch for their calibration.

The probes are calibrated at the beginning and the end of each set of measurements using a fourth-order polynomial and a velocity-pitch map for the SW and XW measurements, respectively.

\subsection{Two-point simultaneous thermal anemometry measurements}\label{sec:twopoint}

For the measurements of the correlation functions for transverse separations presented in  \S \ref{sec:Lu} (datasets 9 -- 11) and of the spanwise derivative components of the velocity presented in \S \ref{sec:Eps} (datasets 14 -- 16), two XW$2.5{\mu}$m are simultaneously used.

\begin{figure}
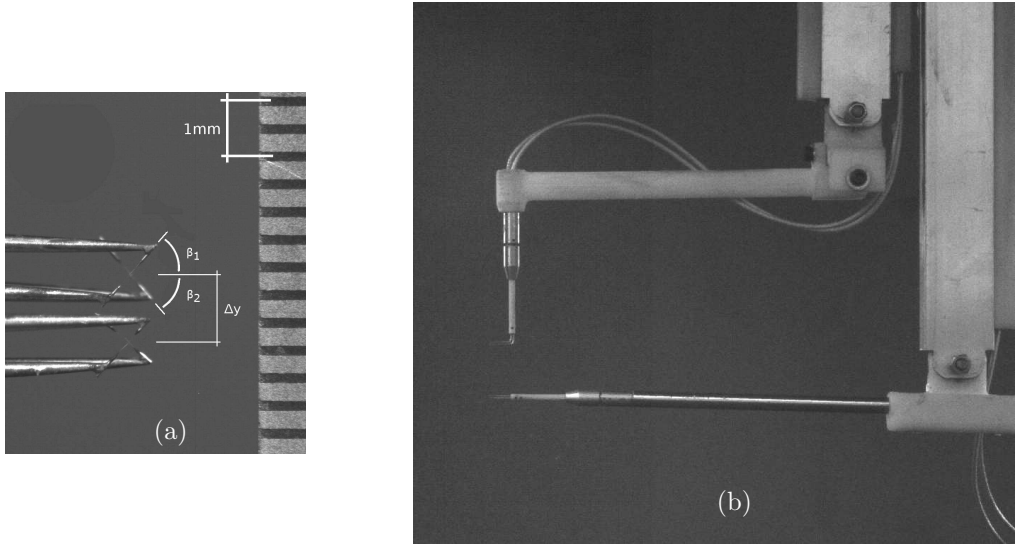

\centering
\begin{minipage}[c]{0.3\linewidth}
   \centering
   \begin{lpic}{Figures/ProbesZoom(40mm)}
   \lbl{160,20;\textcolor{white}{(a)}}
   \end{lpic}
\end{minipage}%
\begin{minipage}[c]{0.6\linewidth}
   \centering 
   \begin{lpic}{Figures/TraverseImage(80mm)}
   \lbl{160,20;\textcolor{white}{(b)}}
   \end{lpic}
\end{minipage}
\caption{Measurement apparatus. (a) Detailed view of the two X-probes and (b) configuration of the apparatus inside the tunnel. }
\label{fig:apparatus}
\end{figure}

The apparatus consists of two X-probes (aligned with the xy plane to measure the longitudinal and vertical velocity components, $U$ and $V$) mounted on a traverse mechanism controlling the vertical distance between the probes and their individual pitch angle for \it in-situ \rm calibration (figure \ref{fig:apparatus}b).
(Note that, in the orthonormal coordinate system used, $x$ is aligned with the mean flow and $y$ \& $z$ are perpendicular and parallel to the floor, respectively.)
The vertical traverse mechanism has single degree of freedom, actuated by a stepper motor, and can only displace the two probes symmetrically about their centroid (defined as the geometrical midpoint between the X-probes' centres; the minimum step is $5 \mu$m). 
Each of the two pitch angle traverses is also actuated by a stepper motor (with a 1:50 ratio gearbox) providing a minimum step angle of $0.018^{\circ}$.
All the stepper motors and gearboxes lie outside of the test section. 
In the case of the  pitch angle traverses the actuation is made via a timing belt running inside each of the two $12\times 25$mm rectangular section tubes (seen in figure \ref{fig:apparatus}b).
The apparatus described above is mounted on a second traverse mechanism controlling the downstream and vertical position of the centroid of the two X-probes and has a minimum traversing step of $40\mu$m and $2.5\mu$m respectively. 

The separation between the two X-probes is measured optically with the aid of an external camera (HiSense 4M camera fitted with a Sigma f/3.5 180mm macro lens and $1.4\times$ tele-converter).
A calibration image is recorded for every set of measurements (i.e. one fixed centroid location and 23 different X-probe separations), such as the one shown in figure \ref{fig:apparatus}a (the original calibration image was cropped and annotated).
The typical field of view and pixel size are $14\times14$mm and $7\mu$m, respectively and the effective focal length is about $250$mm (note that the pixel size is $2-3$ times the wire diameter but it is sufficient to distinguish the wire from the background). 
The location of the centre of each X-probe is inferred from the images as the geometric interception between straight lines   connecting the extremities of the etched portion of the wires.
This differs slightly from the visual interception of the sensors (figure \ref{fig:apparatus}a) since the wires are slightly buckled due to the thermal load they are subjected during operation \cite[]{Perry82} as well as other residual stresses from the soldering/etching process.
The vertical separation between the two X-probes, $\Delta y$, is defined as the vertical distance between the two centres and $\Delta x$ is the downstream separation which should be zero. 
During the course of the experiments it was found that the overall precision of the prescribed vertical separation between the X-probes was typically $\pm 50\mu$m (i.e. over the three degrees of freedom) and the misalignment $\Delta x$ was typically smaller than $200\mu$m (in figure \ref{fig:apparatus}a $\Delta x=50\mu$m).
During the processing of the data the measured X-probes' location ($\Delta y$ is optically confirmed up to $\Delta y=10$mm) is taken into account (the misalignment $\Delta x$ is corrected with the aid of Taylor's hypothesis), even though no noticeable difference was observed when no such corrections are applied. 

The upper and lower X-probes (figure \ref{fig:apparatus}) are modified standard sensors (DANTEC 55P63 and 55P61, respectively) such that the distance between the inclined wires in the z-direction is reduced from $\Delta z \approx 1.0$mm  to $\Delta z \approx 0.5$mm and the original Tungsten wires are replaced by Platinum-(10\%)Rhodium Wollaston wires soldered to the prongs and etched to $l_w\approx 0.45$mm in length and $d_w = 2.5\mu$m in diameter. 
(Note that $\Delta z$ is analogous to $\Delta x_{3}$ in the nomenclature of \cite{ZA96}, see their figure 1.)
The minimum vertical separation between the probes is $\Delta y =1.2$mm (probe resolution $\sim 4\eta - 8\eta$; mutual interference is discussed in \S \ref{sec:resolution}) whereas the maximum separation for datasets 9 -- 11 is $250$mm ($\sim 7 L$) and for  datasets 14 -- 16 is $70$mm ($\sim 2 L$) in a total of $27$ and $23$, respectively, recording locations.
The two X-probes have a small incidence angle relative to the mean flow ($-1^{\circ}$ to $-3^{\circ}$ for upper probe and $1^{\circ}$ to $3^{\circ}$ for lower probe) to guarantee that probes' bodies remain free of contact for small separations.
The angle of the wires relative to the mean flow ($\beta_1$ \& $\beta_2$ in figure \ref{fig:apparatus}a) differ from the standard $\pm 45^{\circ}$ not only due to the incidence angle of the X-probes but also due to the manual soldering of the wires to the prongs. 
For the upper X-probe $\beta_1 = 48^{\circ}$ and $\beta_2=-50^{\circ}$ and for the lower probe $\beta_1 = 48^{\circ}$ and $\beta_2=-41^{\circ}$.
The X-probes are driven by the Streamline CTA system described in \ref{sec:singlepoint} supplemented with two additional channels. 
As for the two-component single-point measurements, the probes are calibrated at the beginning and at the end of each set of measurements using a velocity-pitch map.
At the start of the calibration procedure the X-probes are separated by $\Delta y = 55$mm, which was deemed sufficient to avoid aerodynamic interference between the probes whilst also being sufficiently far from the walls ($>130$mm) at all calibration incidence angles.

For all the measurements, both single-point and two-point, the flow temperature variation from beginning of the first calibration to end of second calibration was less than $1^{\circ}$ thus avoiding the need for temperature corrections to the calibrations \cite[]{Perry82}.

Note also that, for every experimental dataset we ensure that the electronic performance of the CTA system (including the in-build signal conditioners) is sufficient to have an unattenuated response up to frequencies of, at least, $k_1 \eta = 1 $ ($k_1$ is the longitudinal wavenumber and $\eta\equiv (\nu^3/\varepsilon)^{1/4}$ is the Kolmogorov microscale).

\subsubsection{Probe resolution and mutual interference}
\label{sec:resolution}
The common sources of error in the measurement of transverse velocity gradients using two parallel single or X-probes are their finite resolution (individually and of the array), errors in the calibrations of the probes, electronic noise and mutual interference (\citealp{AntoniaBC84}, see also \citealp{Mestayer79,ZA96,ZA02}).  
Errors arising from differences in probe calibrations and electronic noise contamination on each of the probes were found to be negligible when the probe separation is larger than $3\eta$ \cite[pp. 548 of][]{ZA95}, here $\Delta y > 4\eta$. 
Concerning thermal interference due to the proximity of the probes, \cite{AntoniaBC84}  measured quantities like $\overline{(\partial u/\partial x)^2}$ with both probes operating and observed that the quantities remained unchanged when one of the probes was switched off. We repeated the same tests with our X-probes at the minimum separation and corroborate that there is no evidence of thermal interference.  

\begin{figure}
\centering
\includegraphics[width=100mm]{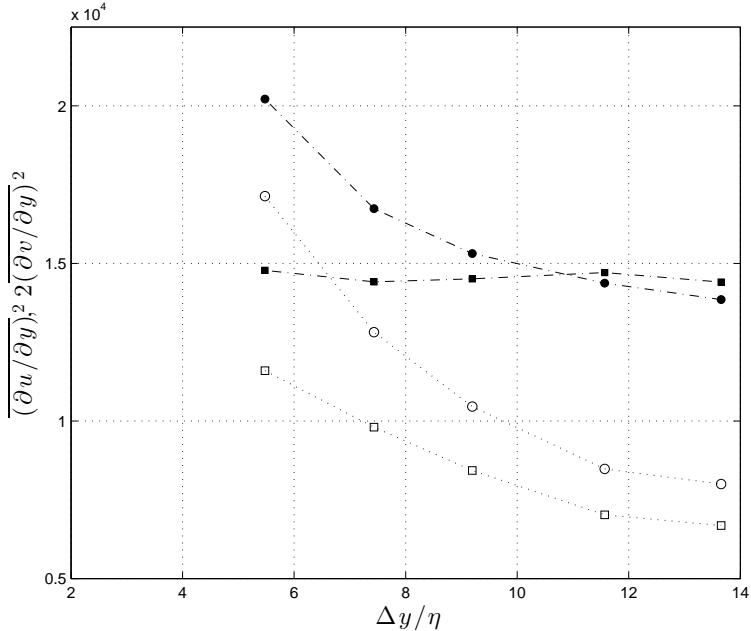}
\caption{Dependence of mean square lateral velocity derivatives (\protect\raisebox{-0.5ex}{\FilledSmallSquare} $\!|\!\!$ \protect\raisebox{-0.5ex}{\SmallSquare}) $\overline{(\partial u/\partial y)^2}$ and (\protect\raisebox{-0.5ex}{\FilledSmallCircle} $\!|\!\!$ \protect\raisebox{-0.5ex}{\SmallCircle}) $2\overline{(\partial v/\partial y)^2}$ on X-probe separation, $\Delta y/\eta$; (empty symbols) raw and (filled symbols) compensated measurements for the attenuation due to finite resolution.}
\label{fig:CorrDissipation}
\end{figure}

The aerodynamic interference due to the proximity of the X-probes depends on the configuration of the measurement apparatus and is investigated with the data from a precursory experiment (measuring RG115-generated turbulence). 
These experiments consisted of traversing, for each downstream location, the X-probes from $\Delta y=1.2$mm to $\Delta y=260$mm with the centroid positioned at (i) $(y,z)=(0,0)$, i.e. the centreline and (ii) $(y,z)=(-57.5\mathrm{mm},0)$, i.e. behind a bar. 
For each downstream location there is a region, $y=-130\mathrm{mm}$ to $73\mathrm{mm}$ which is measured twice.
In particular, the regions around $(y,z)=(0,0)$ and $(y,z)=(-57.5\mathrm{mm},0)$ are measured when the probes are closely spaced, $\Delta y \approx 1.2$mm, and far apart, $\Delta y \approx 115$mm ($2\times 57.5$mm), thus allowing the assessment of aerodynamic interference on single point statistics.
The results show that for $\Delta y < 2$mm the error is never larger than 4\% in quantities like $U$, $u'$, $v'$, $\overline{(\partial u/\partial x)^2}$, $\overline{(\partial v/\partial x)^2}$. 
Higher order statistics such as the Skewness and Kurtosis of both velocity components are less influenced by the X-probes' proximity. 
However, the transverse component of the mean velocity, $V$, is severely influenced by the proximity of the X-probes and for $\Delta y<2$mm errors up to $\pm 0.5\mathrm{ms}^{-1}$ (inlet velocity, $U_{\infty}=10\mathrm{ms}^{-1}$) are observed which may be responsible for the overestimation of the lateral mean square velocity derivative, $\overline{(\partial v/\partial y)^2}$ (see figure \ref{fig:CorrDissipation} and the discussion below).

The spatial resolution of each individual X-probe is always better than $l_w\approx d_w = 3.5\eta$ leading to relatively small finite resolution correction factors to the mean square velocity streamwise derivatives  ($<15\%$).
The correction factors are defined as the ratio between finite resolution and actual values of the 
mean square velocity derivatives ($r_{1,1}\equiv \overline{(\partial u/\partial x)^2}^{\,m}/\overline{(\partial u/\partial x)^2}$ and $r_{2,1}\equiv \overline{(\partial v/\partial x)^2}^{\,m}/\overline{(\partial v/\partial x)^2}$, superscript $m$ indicating measured values, but are omitted throughout the paper). 
In the present study, we use the DNS based correction factors obtained by \cite{B08}.
 
The finite separation between the two X-probes, in addition to the finite resolution of each probe and their aerodynamic interference, influences the estimation of the mean square velocity transverse derivatives, $\overline{(\partial u/\partial y)^2}$ and $\overline{(\partial v/\partial y)^2}$ \cite[]{ZA95,ZA96}. 
The correction factors, $r_{1,2}$($\equiv \overline{(\partial u/\partial y)^2}^{\,m}/\overline{(\partial u/\partial y)^2}$) and $r_{2,2}$ ($\equiv \overline{(\partial v/\partial y)^2}^{\,m}/$ $\overline{(\partial v/\partial y)^2}$), are obtained from figure 3 in \cite{ZA96} to compensate the attenuation due to finite separation (the X-probe geometry for which the correction factors were obtained is not too dissimilar to the one used here).
The influence of the aerodynamic interference of the X-probes in the measurement of the velocity transverse derivatives is assessed by calculating the derivatives for $\Delta y = 1.2, 1.6, 2.0, 2.5$ and $3.0$mm, correcting for the different resolutions and comparing the results. 
If the aerodynamic interference is negligible and the finite separation is correctly compensated, the mean square velocity derivatives should be the same. 
In figure \ref{fig:CorrDissipation}, one example of such a comparison is made for measurements in the lee of the RG115 at  $x=2150$mm.
It is shown that the corrected $\overline{(\partial u/\partial y)^2}$ is indeed roughly independent of the X-probe separation, but $\overline{(\partial v/\partial y)^2}$ is not.
This may be due to the aerodynamic interference already observed in the spanwise mean velocity, $V$. 
Note that the value of $2\overline{(\partial v/\partial y)^2}$ seems to be tending towards the value of $\overline{(\partial u/\partial y)^2}$, which is the expected behaviour for axisymmetric turbulence.

\subsection{Data acquisition and processing} \label{sec:conv}
The in-built signal conditioners of the anemometer are set to analogically filter at  $30$kHz and to offset and amplify the signal $-1$V  and $2\times$, respectively. 
The analogue anemometer signals are sampled at 62.5kHz with a National Instruments NI-6229 (USB) with a resolution of 16-bit over a range of $[-1\,\,1]$V.

The time-varying turbulent signals are converted into spatially-varying by means of Taylor's hypothesis (taking the mean streamwise velocity, $U$, as the advection velocity). 
The signal is digitally filtered at a frequency corresponding to $k_{1} \eta \approx 1.3 $ using a $4^{th}$-order Butterworth filter to eliminate higher frequency noise.
 
The turbulent velocity signals are acquired for $9$min corresponding to $150\,000-200\,000$ integral-time scales. Following \cite{BG96} we quantify the statistical uncertainty of the measurements and conclude that the 95\% confidence interval of the even moments of the longitudinal and transverse velocity components up to fourth-order, as well as the second moment of their derivatives is less than $\pm 1\%$ of the measured value. 

The mean square velocity derivatives in the streamwise direction are estimated from the longitudinal and transverse wavenumber spectra $F_{11}^{(1)}(k_1)$, $F_{22}^{(1)}(k_1)$ as
\begin{equation*}
\overline{(\partial u/\partial x)^2} = \int_{k_{min}}^{k_{max}}k_{1}^2\,F_{11}^{(1)}(k_{1})\, dk_{1}; \,\,\,
\overline{(\partial v/\partial x)^2} = \int_{k_{min}}^{k_{max}}k_{1}^2\,F_{22}^{(1)}(k_{1})\, dk_{1} ,
\end{equation*}
where $k_{min}$ and $k_{max}$ are determined by the window length and the sampling frequency respectively. To negotiate the problem of low signal to noise ratios at high frequencies we follow \cite{antoniadissipation} and fit an exponential curve to the high frequency end of the spectra. We checked that this does not change the mean square velocity derivatives by more than a few percent. However, in decaying turbulent flows the signal to noise ratio progressively decreases as the flow decays and therefore it is preferred to remove any high frequency noise to avoid a systematic bias to the data, regardless of how small this bias may be.


\section{Profiles of one-point turbulence statistics} \label{sec:topology}

\subsection{Production and decay regions} \label{sec:wake}
The downstream evolution of the turbulent flow generated by the RGs and the FSGs can be separated into two distinct regions, the production and the decay regions. 
The production region lies in the immediate vicinity downstream of the grid where individual wakes generated by individual bars develop and interact. 
This region extends as far downstream as where the wakes of the biggest bars interact, i.e. as far as that distance downstream where the width of these largest wakes is comparable to the largest mesh size $M$ \cite[][$M$ is the distance between parallel bars within a mesh]{MV2010}. 
There is only one mesh size in the case of RGs but for FSGs, which are made of many different square bar arrangements, i.e. meshes, of different sizes, $M$ refers to the largest mesh size (see figure \ref{fig:grids} and table \ref{table:grids}).

In the case of our RGs and of the particular type of space-filling low-blockage FSGs with high enough thickness ratio, $t_r$ (such as the ones studied here and previously by \cite{SV2007}, \cite{MV2010} and \cite{VV2011,VV2012}; for the definition of $t_r$ see caption of table \ref{table:grids}) the turbulent kinetic energy in the production region increases monotonically with downstream distance $x$ along the centreline until it reaches a maximum
at $x=x_{\mathrm{peak}}$ (see figure \ref{fig:UmvsX}). 
Further downstream, i.e. where $x > x_{\mathrm{peak}}$, the turbulence decays monotonically. 
Along any other line parallel to the centreline the point downstream beyond which the turbulence decays monotonically occurs before $x_{\mathrm{peak}}$ as shown by \cite{JW1992} and \cite{Ertunc2010} for RGs and by \cite{MV2010} and \cite{Sylvain2011} for FSGs. 
Note that this definition of production and decay separated by a plane perpendicular to the centreline located at  $x_{\mathrm{peak}}$ could in principle be made more precise by defining the surface where the advection vanishes, $U_{k}\partial \overline{q^2}/\partial x_k = 0$ (i.e. where the turbulent kinetic energy, $\overline{q^2}/2$ is maximum; $\overline{q^2}\equiv \overline{u_i\,u_i}$).

In the present paper only the region downstream of $x_{\mathrm{peak}}$, up to a distance $x$ from the grid equal to the first few multiples of $x_{\mathrm{peak}}$, is investigated.
This is the region where the nonequilibrium dissipation behaviour was reported by \cite{SV2007}, \cite{MV2010}, \cite{VV2011}, \cite{gomesfernandesetal12} and \cite{discettietal11} for the FSGs and \cite{VV2012} for the RGs. 
The overall extent of this region remains unknown except for the RG60 grid where the downstream extent of the
test section and measurements were sufficient to capture the existence of an equilibrium region beyond the nonequilibrium one; see \cite{VV2012} who determined the cross-over distance between these two regions to be about $5x_{\mathrm{peak}}$ from the RG60 grid.
Note that the downstream location of   $x_{\mathrm{peak}}$ was shown to be proportional to a `wake-interaction length-scale', $x_* \equiv M^2/t_0$ \cite[see][]{MV2010,VV2011}, and the proportionality constant to be dependent on other grid details as well as upstream turbulence \cite[]{gomesfernandesetal12}.

Before proceeding with the profiles of one-point turbulence statistics, it may now be a good place to explain why the
non-equilibrium dissipation region had never been reported before the advent of fractal grids in 2007. 
Regular grids had been designed to have as high a blockage ratio ($\sigma$) as possible in order for the
turbulence intensities and local Reynolds numbers to be as high as possible. 
However, the ensuing flow is dominated by instabilities and wake mergings if $\sigma$ is too high \cite[see][]{CorrsinHandbook} and so the vast majority of regular grids used since the 1930s when they started
being used for turbulence research had a $\sigma$ between about 35\% and 45\%. 
There is a one-to-one decreasing relationship between $M/t_0$ and $\sigma$ for regular grids which is such that $M/t_0 \to
\infty$ as $\sigma \to 0$ and $M/t_0 = 1$ at $\sigma =1$ (i.e. 100\%). 
Indicatively, for $\sigma=$ 40\% one gets $M/t_{0}\approx 4.5$ and therefore $x_{*} \approx 4.5M$ giving an estimate for
$x_{peak} \approx 2.2M$ \cite[see][]{MV2010,VV2011,gomesfernandesetal12} and an estimate of the extent of the non-equilibrium region to be from $2.2M$ to $11M$ (assuming the estimate of \citealt{VV2012} for RG60 in terms of multiples of $x_{peak}$ to carry over to other RGs). 
As the safe application of hot-wire anemometry in conjunction with the Taylor frozen turbulence hypothesis requires turbulence intensities not exceeding about 15\%, the vast majority of measurements taken with RGs over the decades where made at distances from the grid at and beyond $10M$, hence missing the non-equilibrium dissipation region
completely. 
Furthermore, the ratio between the tunnel width and the mesh size has very often been around 10 or even higher, particularly when attempts were made to simulate homogeneous turbulence. 
Hence the vast majority of the tunnel test section could and would be used to probe distances from the grid well over $x/M \approx 20$.

The constraints in designing fractal grids encountered by \cite{HV2007} led them to adopt much lower blockage ratios than
usual, specifically $\sigma = $ 25\% for FSGs, see table \ref{table:grids}. 
As a result, $M/t_{0} = 12$ for FSG3'x3' and FSG18''x18'' suggesting that the non-equilibrium dissipation region extends from about $x_{peak} \approx 0.4 M (M/t_{0}) \approx 5M$ to about $5x_{peak} \approx 24M$ and therefore well beyond the end of the test section. 
The blockage ratio being small, turbulence intensities are limited to values below about 10\% in the non-equilibrium dissipation region and therefore hot wire anemometry can be used to study it in conjunction with the Taylor frozen turbulence hypothesis. 
An added advantage is that, throughout the test section beyond $x_{peak}$, the turbulence intensity does not drop down to the very low values encountered with typical regular grids where the emphasis is given on measurements at
high $x/M$, thus avoiding signal-to-noise problems. 
Hence FSGs paved the way for turbulence experiments where the non-equilibrium dissipation region is magnified and covers the entire extent of the test section beyond $x_{peak}$, yet the turbulence intensities are such that hot wire anemometry can be used conclusively.

\cite{VV2012} followed by designing the untypical regular grids RG230 and RG115 of figure \ref{fig:grids} which also have a low
blockage ratio and a high $M/t_0$ and which also return long non-equilibrium dissipation regions with similar turbulence 
intensities making hot wire anemometry similarly suitable. 
Hence, \cite{VV2012}  demonstrated that the non-equilibrium dissipation region is not unique to fractal grids. 
Their two low-blockage RGs (RG230 and RG115) and the two FSGs of figure \ref{fig:grids} and table \ref{table:grids} are used in the remainder of section 3 to document the differences and similarities in the non-equilibrium dissipation regions of the turbulent flows they generate and the potential effects of the confining walls of the tunnel. 
Thus we demonstrate that the new non-equilibrium dissipation law which we study further in sections 4 and 5 holds irrespective of some significant differences in profiles and wall-confinement effects. 
This demonstration would not have been possible without the FSGs and the new RGs.

\subsection{Mean velocity deficit versus turbulent kinetic energy decay} \label{sec:profiles}
\begin{figure}
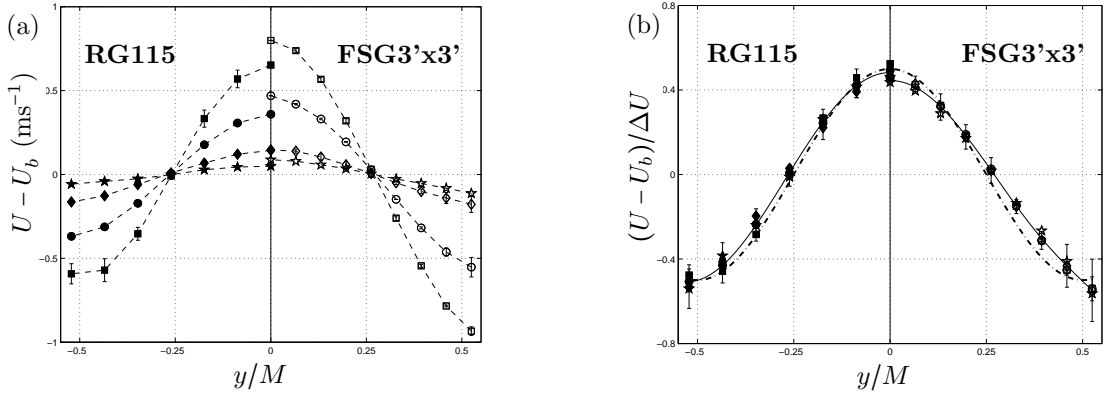

\begin{minipage}[c]{0.5\linewidth}
   \centering
   \begin{lpic}{Figures/UmMinusUbProfileRG115FSG3x3(65mm)}
   \lbl{10,130;(a)}
   \lbl{50,120;\bf{RG115}}
   \lbl{150,120;\bf{FSG3'x3'}}
   \lbl[W]{100,0;$y/M$}
   \lbl[W]{0,85;${}^{}_{}$\hspace{15mm}}         
   \lbl[W]{10,85,90;$U-U_b^{}$  (ms$^{-1}$)}         
   \end{lpic}
\end{minipage}%
\begin{minipage}[c]{0.5\linewidth}
   \centering 
   \begin{lpic}{Figures/UmMinusUbOverDeltaUProfileRG115FSG3x3(65mm)}
   \lbl{10,130;(b)}
   \lbl{50,120;\bf{RG115}}
   \lbl{150,120;\bf{FSG3'x3'}}
   \lbl[W]{100,0;$y/M$}
   \lbl[W]{9,80,90;$(U-U_b)/\Delta U^{}$}      
   \end{lpic}
\end{minipage}
\caption{Mean velocity transverse profiles (normalised by the mean velocity deficit, $\Delta U$$\equiv U(y=0)-U(y=\pm M/2)$ in (b)), for different downstream locations in the lee of the RG (filled symbols) and the FSG (empty symbols). The bulk velocity, $U_b$ (see text for definition) is subtracted from the velocity profiles.
Downstream locations: (\protect\raisebox{-0.5ex}{\FilledSmallSquare} $\!|\!\!$ \protect\raisebox{-0.5ex}{\SmallSquare}) $x/x_{\mathrm{peak}}=1.4,\,1.5$; (\protect\raisebox{-0.5ex}{\FilledSmallCircle} $\!|\!\!$ \protect\raisebox{-0.5ex}{\SmallCircle}) $x/x_{\mathrm{peak}}=1.8,\,2.0$; (\protect\raisebox{-0.5ex}{\FilledDiamondshape} $\!|\!\!$ \protect\raisebox{-0.5ex}{\Diamondshape}) $x/x_{\mathrm{peak}}=2.8,\,3.0$; (\ding{72} $|\!\!$ \ding{73}) $x/x_{\mathrm{peak}}=3.7,\,3.5$.  
All data are recorded at $U_{\infty}=15$ms$^{-1}$.
The dash-dotted line in figure (b) represents a cosine law, $cos(\theta)/2$ with $\theta = [-\pi\,\,\,\,\pi]$ corresponding to $y =[-\mathrm{M}/2\,\,\,\,\mathrm{M}/2]$ and the solid lines represent a $6^{\mathrm{th}}$-order polynomial fit.
Error bars represent the departures from symmetry between the upper ($y/M>0$) and lower ($y/M<0$) half of the transverse measurements whereas the symbols represent the mean between the two.
All data are taken in the plane $z=0$ (datasets 2 and 6 are used, see table \ref{table:summary}).}
\label{fig:Um}
\end{figure}
\begin{figure}
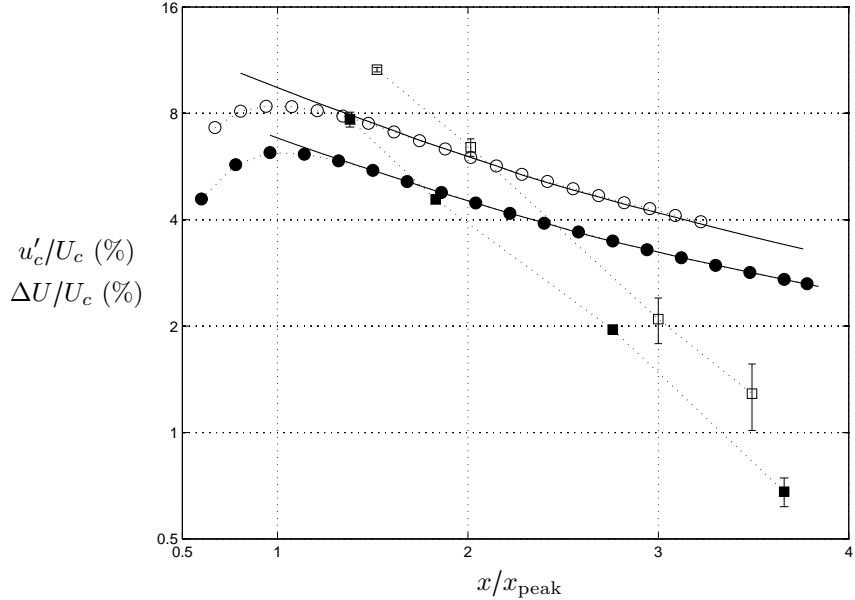

\centering
   \begin{lpic}{Figures/DeltaUvsTuLongitudinalEvo(110mm)}
   \lbl[W]{116,2;$x/x_{\mathrm{peak}}$}
   \lbl[W]{10,74;${}_{}^{}$\hspace{20mm}}      
   \lbl[W]{12,80;$u'_c/U_c$ (\%)}      
   \lbl[W]{12,70;$\Delta U/U_c$ (\%)}      
   \end{lpic}
\caption{Downstream evolution of (\protect\raisebox{-0.5ex}{\FilledSmallCircle} $\!|\!\!$ \protect\raisebox{-0.5ex}{\SmallCircle}) turbulence intensity at centreline, $u'_c/U_c$ and (\protect\raisebox{-0.5ex}{\FilledSmallSquare} $\!|\!\!$ \protect\raisebox{-0.5ex}{\SmallSquare}) mean velocity deficit normalised by the centreline velocity, $\Delta U/U_c$ in the lee of the RG115 (filled symbols) and the FSG3'x3' (empty symbols). 
The solid lines are power-law fits to the decay of $u'_c/U_c$ using the nonlinear least-squares regression method described in \cite{VV2011} (`Method III' in their \S 3.4). The decay exponents and virtual origins obtained are $n=3.0,\, 2.4$ and $x_0/x_{\mathrm{peak}} = -1.4,\,-1.1$ for the FSG3'x3' and RG115 data, respectively. Datasets 1, 2,  6 and 7 are used, see table \ref{table:summary}. }
\label{fig:UmvsX}
\end{figure}

We start by examining the mean velocity profiles as the turbulent flow decays and compare, in figure
\ref{fig:Um}, those resulting from the RG115 and the FSG3'x3' grids.
Note that the bulk velocity, $U_b\equiv 1/M \int_{-M/2}^{M/2}$ \!\! U dy, is subtracted from the mean velocity  profiles to compensate for the slight decrease in the effective area of the test section due to the blockage caused by the developing boundary-layers on the side-walls. 
(For wind tunnels with mechanisms to compensate boundary-layer growth, e.g. a divergence test section, $U_b=U_{\infty}$.)
Normalising the profiles with the velocity deficit, $\Delta U$  (defined in the caption of figure \ref{fig:Um}), it can be seen that the mean profiles retain approximately the same shape as the velocity deficit decreases (figure \ref{fig:Um}b). 
In the RG115-generated turbulence case this profile is not very dissimilar from a cosine law.
However it does seem to have some slight deviations from the cosine law in the FSG3'x3'-generated turbulence case which must be attributable to the change in upstream conditions, i.e. grid geometry.

The decay of $\Delta U$ is faster than the decay of $u'$ (the root mean square (rms) of the longitudinal component of the fluctuating velocity), see figure \ref{fig:UmvsX}. 
This starkly differs from a wake-like flow where scaling arguments suggest that $u' \sim \Delta U$ \cite[]{TennekesLumley:book}. 

\subsection{Profiles of $2^{\mathrm{nd}}$-order one-point turbulence statistics}\label{sec:2nd}
\begin{figure}
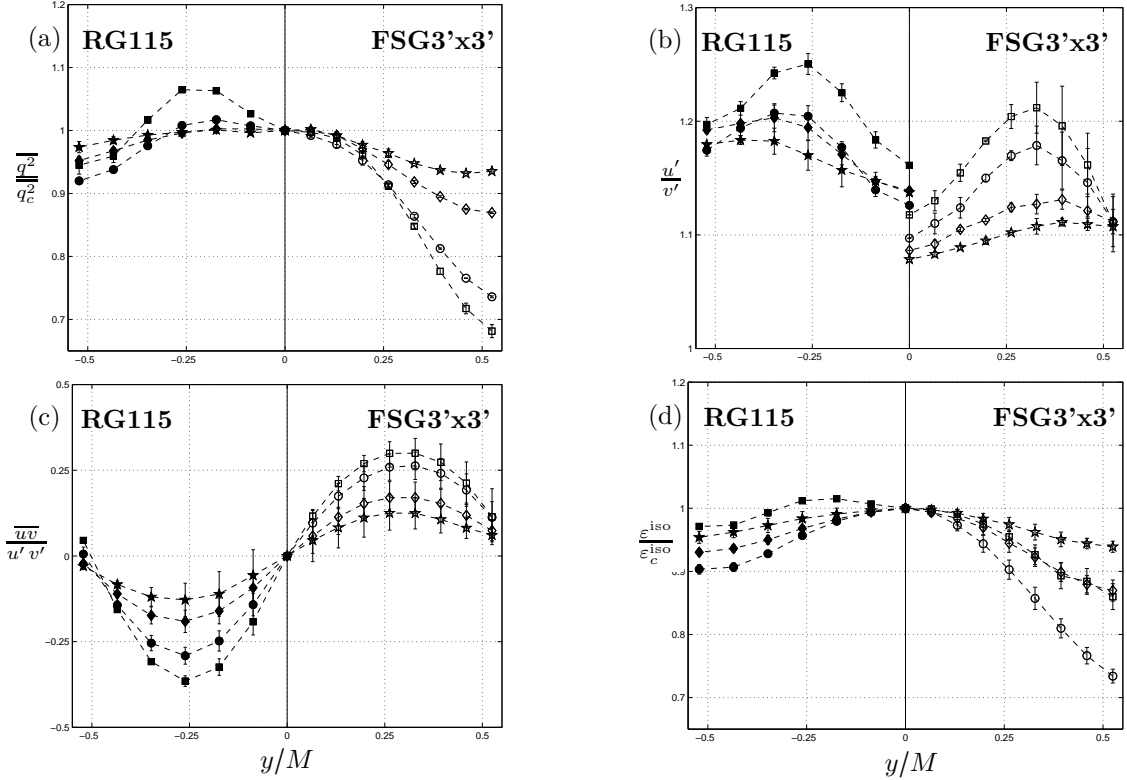

\centering
\begin{minipage}[c]{0.5\linewidth}
   \centering
   \begin{lpic}[b(-2mm)]{Figures/q2ProfileRG115FSG3x3(65mm)}
   \lbl{10,125;(a)}
   \lbl{40,125;\bf{RG115}}
   \lbl{150,125;\bf{FSG3'x3'}}
   \lbl[W]{5,75;$\frac{\overline{q^2}}{\overline{q_c^2}}$}        
   \lbl[W]{100,3;${}^{}$\hspace{5mm}} 
   \end{lpic}
\end{minipage}%
\begin{minipage}[c]{0.5\linewidth}
   \centering 
   \begin{lpic}[b(-2mm)]{Figures/uovervProfile(65mm)}
   \lbl{10,125;(b)}
   \lbl{40,125;\bf{RG115}}
   \lbl{150,125;\bf{FSG3'x3'}}
   \lbl[W]{11,75;\hspace{2mm}$\frac{u'}{v'}$}        
   \lbl[W]{100,3;${}^{}$\hspace{5mm}}    
   \end{lpic}
\end{minipage}
\begin{minipage}[c]{0.5\linewidth}
   \centering
   \begin{lpic}{Figures/uvProfile(65mm)}
   \lbl{10,125;(c)}
   \lbl{40,125;\bf{RG115}}
   \lbl{150,125;\bf{FSG3'x3'}}
   \lbl[W]{100,0;$y/M$}
   \lbl[W]{5,80;$\frac{\overline{uv}}{u'\,v'}$}           
   \end{lpic}
\end{minipage}%
\begin{minipage}[c]{0.5\linewidth}
   \centering 
   \begin{lpic}{Figures/EpsProfile(65mm)}
   \lbl{10,125;(d)}
   \lbl{40,125;\bf{RG115}}
   \lbl{150,125;\bf{FSG3'x3'}}
   \lbl[W]{100,0;$y/M$}
   \lbl[W]{5,70;${}^{}$}           
   \lbl[W]{8,80;$\frac{\varepsilon^{\mathrm{iso}}}{\varepsilon_c^{\mathrm{iso}}}$}           
   \end{lpic}
\end{minipage}
\caption{Reynolds stress transverse profiles for different downstream locations in the lee of RG115 and FSG3'x3'. Symbols  and error bars are described in the caption of figure \ref{fig:Um}. All data are taken in the plane $z=0$. Datasets 2 and 6 (see table \ref{table:summary}) are used as in figure \ref{fig:Um}.}
\label{fig:ReyStress}
\end{figure}
\begin{figure}
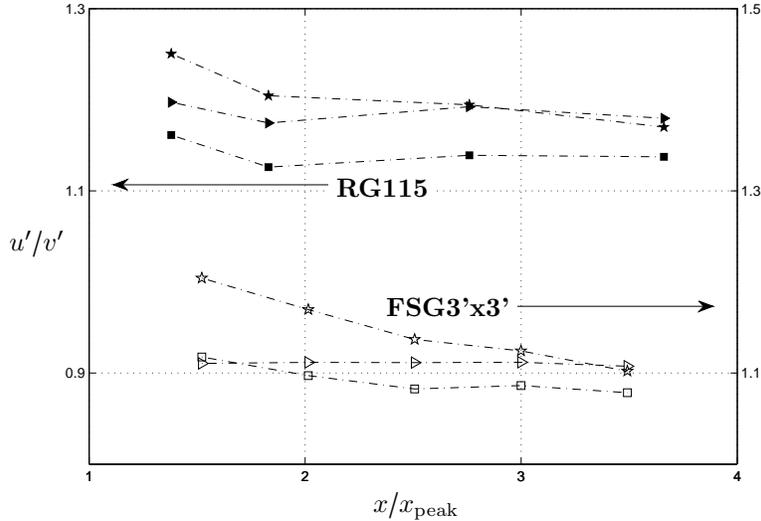

\centering
   \begin{lpic}{Figures/uovLongProfileRGFG(100mm)}
   \lbl[W]{102,2;$x/x_{\mathrm{peak}}$}
   \lbl[W]{5,75;\hspace{2mm}${}^{}$}        
   \lbl[W]{5,65;\hspace{2mm}${}^{}$}        
   \lbl[W]{5,68;\hspace{2mm}$u'/v'$}        
   \lbl[W]{93.5,82;\bf{RG115}}
   \lbl[W]{110,52;\bf{FSG3'x3'}}
   \end{lpic}
   \caption{Downstream evolution of the ratio between the longitudinal and spanwise rms velocities at  
(\protect\raisebox{-0.5ex}{\FilledSmallSquare} $\!|\!\!$ \protect\raisebox{-0.5ex}{\SmallSquare}) $y=0$, 
(\ding{72} $|\!\!$ \ding{73}) $y=-M/4$ and 
(\protect\raisebox{-0.5ex}{\FilledSmallTriangleRight} $\!|\!\!$ \protect\raisebox{-0.5ex}{\SmallTriangleRight}) $y=-M/2$ 
in the lee of the RG115 (filled symbols) and the FSG3'x3' (empty symbols). Datasets 2 and 6 (see table \ref{table:summary}) are used as in figure \ref{fig:Um}}
\label{fig:uov}
\end{figure}


We now turn to transverse profiles of the turbulent kinetic energy $\overline{q^{2}}/2 = \overline{u^{2}}/2 +\overline{v^{2}}/2 +\overline{w^{2}}/2 $ estimated here as $\overline{u^{2}}/2 + \overline{v^{2}}$. 
This estimate relies on the assumption that $\overline{v^{2}} \approx \overline{w^{2}}$ in the decay region $x >
x_{\mathrm{peak}}$ which is supported as a rough approximation by the direct numerical simulations of \cite{Sylvain2011} and the laboratory experiments of \cite{Nagata2012}.

Our first observation is that the shape of the profile changes and becomes progressively more uniform as the turbulent flow decays (figure \ref{fig:ReyStress}a).
(The profiles are normalised by the centreline value to enhance variation of the kinetic energy with spanwise location.)
This is indeed what one expects of grid-generated turbulence which has a tendency towards asymptotical homogeneity, at least in some respects, with downstream distance.
A striking difference, however, is that the profiles of the RG115-generated turbulence are more uniform than their counterparts of the FSG3'x3'-generated turbulence at similar downstream locations relative to $x_{\mathrm{peak}}$. 
This may be attributed to the additional turbulence generated by the wakes originating from the smaller squares of the FSG near the centreline (compare figure \ref{fig:grids}a with figure \ref{fig:grids}d near the centre of the grid), which increase  the kinetic energy in this region. 
A similar effect appears in the profiles of the isotropic dissipation estimate $\varepsilon^{\mathrm{iso}} = 15 \nu \overline{(\partial u/ \partial x)^{2}}$ which we plot in figure \ref{fig:ReyStress}d (normalised by the values of $\varepsilon^{\mathrm{iso}}$ on the centreline, $y=z=0$).
(Note, however, that the transverse profiles of the actual kinetic energy dissipation rate $\varepsilon$ divided by $\varepsilon^{\mathrm{iso}} $ may be different in the different turbulent flows considered here.)
Even though the present FSGs return less homogeneity than the present RGs, they do nevertheless seem to generate a significant improvement in the $u'/v'$ ratio (figure \ref{fig:ReyStress}b) which is one of the indicators of large-scale
isotropy ($v'$ is the rms of the turbulent fluctuating velocity in the $y$ direction).
The downstream evolution of this ratio is presented in figure \ref{fig:uov} where it can also be seen that $u'/v'$ do not vary significantly during decay \cite[in-line with the literature, see \S 3.9 in][]{Townsend:book}.

Note that the curvature of the kinetic energy transverse profiles is associated with the lateral triple correlation transport, $\partial /\partial y \, \overline{vq^2}$.
This can be seen from an eddy diffusivity estimate $\overline{vq^2} = -\mathrm{D_T}\, \partial /\partial y \, \overline{q^2}$
which leads to $\partial /\partial y \, \overline{vq^2} = -\mathrm{D_T}\, \partial^2 /\partial y^2 \, \overline{q^2}$ if the eddy diffusivity $\mathrm{D_T}$ is independent of $y$.
Profiles of triple correlation transport terms are presented in \S \ref{sec:homo} showing opposite net transport near the centreline at a distance of about $1.5x_{\mathrm{peak}}$ for the two grids (RG115 and FSG3'x3').
The opposite curvatures appearing near the centreline in the $\overline{q^{2}}$ transverse profiles at that distance from each grid (figure \ref{fig:ReyStress}a) directly relate to the opposite signs of the lateral triple correlation transports generated by the two grids at these locations.

Figure \ref{fig:ReyStress}c shows that differences in the Reynolds shear stress (normalised by the local $u'$ and $v'$) between the turbulence generated by the two grids are very tenuous. 
The numerical values of the normalised shear stress are similar and the shape of the profiles differs only slightly in the location of the peak and the numerical values at $y=M/2$.

\subsection{Wind tunnel confinement effects} \label{sec:confinement}
\begin{figure}
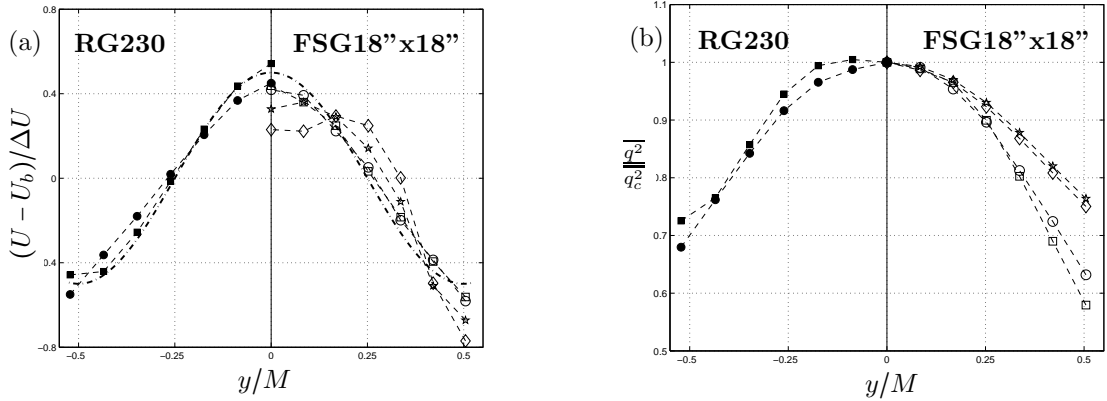

\centering
\begin{minipage}[c]{0.5\linewidth}
   \centering
   \begin{lpic}{Figures/UmMinusUbOverDeltaUProfileRG230FSG18(65mm)}
   \lbl{10,125;(a)}
   \lbl{45,125;\bf{RG230}}
   \lbl{140,125;\bf{FSG18''x18''}}
   \lbl[W]{100,0;$y/M$}
   \lbl[W]{9,75,90;$(U-U_b)/\Delta U^{}$}         
   \end{lpic}
\end{minipage}%
\begin{minipage}[c]{0.5\linewidth}
   \centering 
   \begin{lpic}{Figures/q2ProfileRG230FSG18(65mm)}
   \lbl{10,125;(b)}
   \lbl{45,125;\bf{RG230}}
   \lbl{140,125;\bf{FSG18''x18''}}
   \lbl[W]{7,80;$\frac{\overline{q^2}}{\overline{q_c^2}}$}        
   \lbl[W]{100,3;${}^{}$\hspace{5mm}} 
   \lbl[W]{100,0;$y/M$}    
   \end{lpic}
\end{minipage}%
\caption{Transverse profiles of (a) the normalised mean velocity and (b) kinetic energy for different downstream locations in the lee of the RG230 (filled symbols) and the FSG18''x18'' (empty symbols). Downstream locations: (\protect\raisebox{-0.5ex}{\FilledSmallSquare} $\!|\!\!$ \protect\raisebox{-0.5ex}{\SmallSquare}) $x/x_{\mathrm{peak}}=1.3,\,1.4$; (\protect\raisebox{-0.5ex}{\FilledSmallCircle} $\!|\!\!$ \protect\raisebox{-0.5ex}{\SmallCircle}) $x/x_{\mathrm{peak}}=1.8,\,1.8$; (\protect\raisebox{-0.5ex}{\Diamondshape}) $x/x_{\mathrm{peak}}=2.8$; (\ding{73}) $x/x_{\mathrm{peak}}=3.2$.  
The dash-dotted line on figure (a) represents a cosine law. Datasets 3 and 5 (see table \ref{table:summary}) are used.}
\label{fig:ReyStressB}
\end{figure}

In \cite{VV2011} (\S 3.2.3) some concerns were raised regarding the effect of the wind tunnel bounding walls on the measured turbulent flow. 
As noted by \cite{HG1978} the solid walls have a blocking action on the large-scale free-stream turbulence eddies adjacent to the wall up to a distance of the order of the integral-length scale (even in the absence of mean shear for the case of moving walls with a tangential velocity equal to that of the mean flow). 
\cite{VV2011} compared the ratio of their wind tunnel width/height and their integral length-scales (estimated to range between 8-10) with DNS and other grid-turbulence experiments and argued that the effect of confinement could not, by itself, justify the outstanding properties of their flow, in particular the constancy during decay of the integral to Taylor length-scale ratio $L_{11}^{(1)}/\lambda$ (see \S \ref{sec:Lu} for the definition of this notation). 
Nonetheless, there could be some effect on that ratio, for example by artificially reducing it and causing it to slightly decrease rather than remain constant during decay.

The effect of wind tunnel confinement is investigated here by comparing geometrically similar grids with different ratios between test section width/height ($T$) and mesh size ($M$).
This is accomplished by (i) comparing mean profiles from the FSG18''x18'' and the FSG3'x3' grid arrangements for which
$M$ is approximately the same but the FSG3'x3' is a periodic extension of the FSG18''x18'' in a wind tunnel of double the size; and (ii) by comparing mean profiles from the RG230 and RG115 arrangements which are geometrically similar in the same wind tunnel but where the mesh sizes differ by a factor 2. 
These two comparisons provide an assessment of (i) the effect of generating large integral length-scales relative to the tunnel's cross section (which could influence, e.g., the downstream evolution of the $L_{11}^{(1)}/\lambda$ ratio, see \S \ref{sec:Llambda}) and (ii)  the difference between wakes interacting with each other whilst simultaneously interacting with the wall (as is the case for RG230- and FSG18''x18''-generated turbulence) versus wakes interacting with each other in a quasi-periodic arrangement  (as is the case in the centre regions of the RG115- and FSG3'x3'-generated turbulent flows).

Comparing the normalised mean velocity profiles of RG230 and FSG18''x18'' in figure \ref{fig:ReyStressB}a with those of RG115 and FSG3'x3' in figure \ref{fig:Um}b, it is clear that the profiles corresponding to the grids with double the value of
$M/T$ have lost the similarity with downstream position which characterises the profiles resulting from the lower $M/T$ grids. 
This effect is more pronounced further downstream, in particular at the furthermost downstream stations where the greatest departures from self-similar profile shape appears (see figure \ref{fig:ReyStressB}a). 
The two furthermost stations in the FSG18''x18'' case are at $x=3650$mm and $x=4250$mm whereas the last
measurement station in the RG230 case is $x=3050$mm. 
It is therefore no surprise that the greatest deviations from self-similar mean flow profile are evidenced in the FSG18''x18'' case as the blockage induced by the boundary layers developing on the confining walls is greater at $x=3650$mm and $x=4250$mm than at $x=3050$mm.

Turning to the effect on the kinetic energy profiles (figure \ref{fig:ReyStressB}b versus figure \ref{fig:ReyStress}a) it can be seen that there is a substantial decrease in the uniformity of the profiles across the transverse locations for the grids with higher $M/T$ (RG230 and FSG18''x18''). 
It also appears that this effect is felt throughout the decay and is as pronounced closer to $x_{\mathrm{peak}}$ as it is further downstream.
For example, the overshoot of kinetic energy off the centreline observed at the first measurement station ($x/x_{\mathrm{peak}}=1.4$) in the lee of RG115 (c.f. figure \ref{fig:ReyStress}a) is almost non-existent for the RG230 data at an identical downstream location relative to $x_{\mathrm{peak}}$ ($x/x_{\mathrm{peak}}=1.3$, c.f. figure \ref{fig:ReyStressB}b).
It is likely that this difference is a consequence of the wakes generated by the RG230 and FSG18''x18'' grids interacting with the wall instead of interacting with other wakes from a periodic bar, as in the case of RG115 and FSG3'x3'. 
These changes in the shape of the profiles lead to changes in the turbulent transport as shown in \S \ref{sec:homo}.

\subsection{Turbulent transport and production} \label{sec:homo}

\begin{figure}
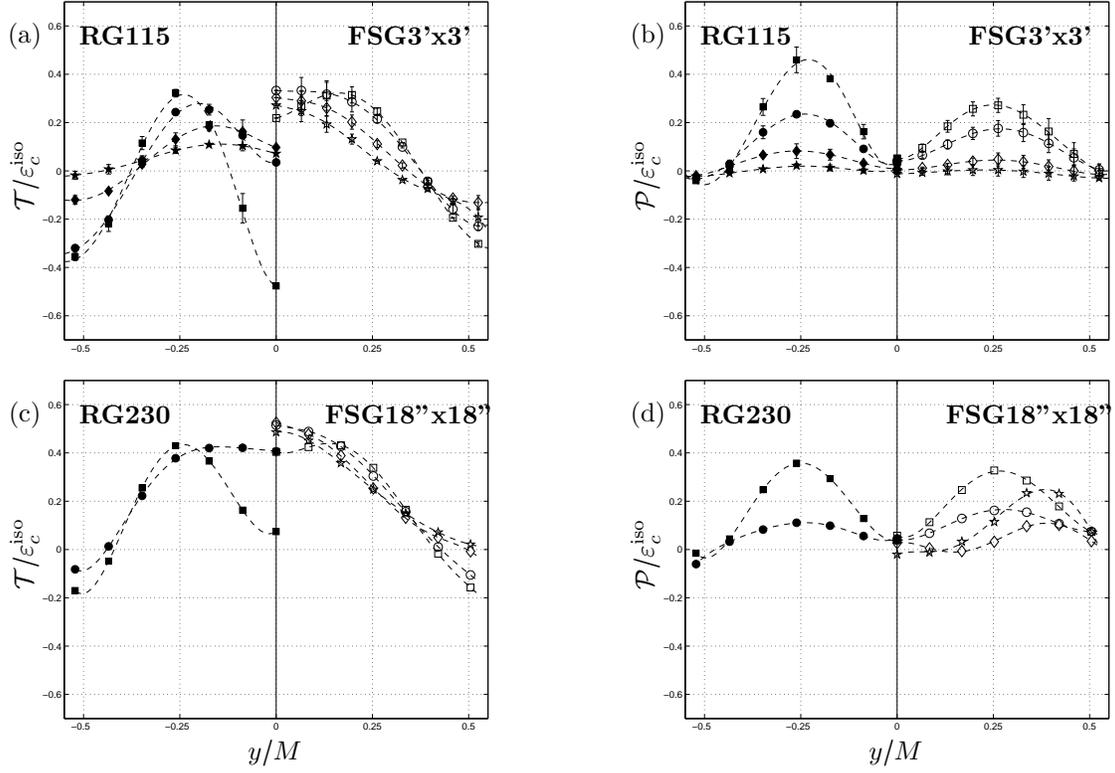

\centering
\begin{minipage}[c]{0.5\linewidth}
   \centering
   \begin{lpic}{Figures/TranspTransvProfile(65mm)}
   \lbl{8,125;(a)}
   \lbl{45,125;\bf{RG115}}
   \lbl{150,125;\bf{FSG3'x3'}}
   \lbl[W]{90,3;${}^{}$\hspace{15mm}}
   \lbl[W]{8,75,90;$\mathcal{T}/\varepsilon_c^{\mathrm{iso}}$}   
   \end{lpic}
\end{minipage}%
\begin{minipage}[c]{0.5\linewidth}
   \centering 
   \begin{lpic}{Figures/ProdTransvProfile(65mm)}
   \lbl{9,125;(b)}
   \lbl{45,125;\bf{RG115}}
   \lbl{150,125;\bf{FSG3'x3'}}
   \lbl[W]{90,3;${}^{}$\hspace{15mm}}
   \lbl[W]{8,75,90;$\mathcal{P}/\varepsilon_c^{\mathrm{iso}}$}
   \end{lpic}
\end{minipage}
\begin{minipage}[c]{0.5\linewidth}
   \centering
   \begin{lpic}{Figures/TranspTransvProfileRG230FSG18(65mm)}
   \lbl{8,125;(c)}
   \lbl{45,125;\bf{RG230}}
   \lbl{150,125;\bf{FSG18''x18''}}
   \lbl[W]{102,0;$y/M$}   
   \lbl[W]{8,75,90;$\mathcal{T}/\varepsilon_c^{\mathrm{iso}}$}   
   \end{lpic}
\end{minipage}%
\begin{minipage}[c]{0.5\linewidth}
   \centering 
   \begin{lpic}{Figures/ProdTransvProfileRG230FSG18(65mm)}
   \lbl{9,125;(d)}
   \lbl{45,125;\bf{RG230}}
   \lbl{150,125;\bf{FSG18''x18''}}
   \lbl[W]{102,0;$y/M$}   
   \lbl[W]{8,75,90;$\mathcal{P}/\varepsilon_c^{\mathrm{iso}}$}
   \end{lpic}
\end{minipage}%
\caption{Transverse profiles of turbulent (a, c) transport and (b, d) production for different downstream locations in the lee of the various RGs and FSGs. Symbols and error bars for the top plots (a, b) are described in the caption of figure \ref{fig:Um} and  symbols for the bottom plots (c, d) in figure \ref{fig:ReyStressB}. Datasets 2, 3, 5 and 6 (see table \ref{table:summary}) are used.}
\label{fig:TKESpanProf}
\end{figure}

\begin{figure}
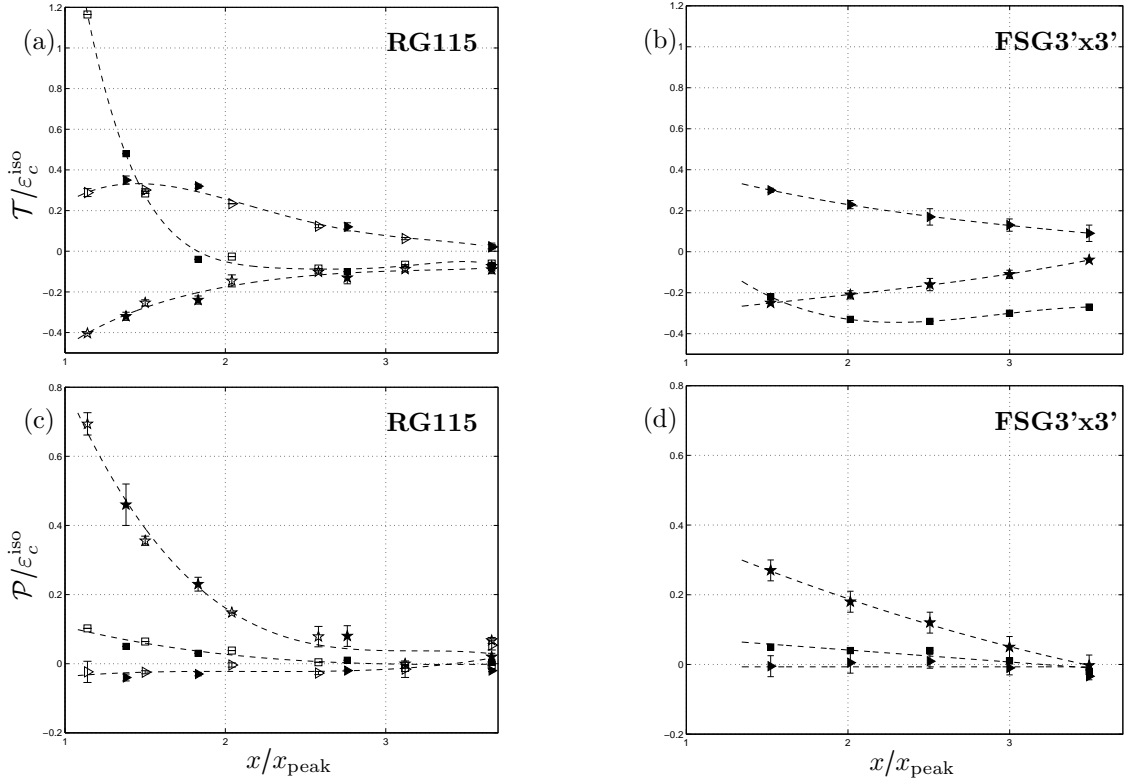

\centering
\begin{minipage}[c]{0.5\linewidth}
   \centering
   \begin{lpic}[b(-2mm)]{Figures/TranspLongProfileRG(65mm)}
   \lbl{8,125;(a)}
   \lbl{150,125;\bf{RG115}}
   \lbl[W]{90,2;${}^{}$\hspace{15mm}}
   \lbl[W]{2,75,90;$\mathcal{T}/\varepsilon_c^{\mathrm{iso}}$}
   \end{lpic}
\end{minipage}%
\begin{minipage}[c]{0.5\linewidth}
   \centering 
   \begin{lpic}[b(-2mm)]{Figures/TranspLongProfileFG(65mm)}
   \lbl{8,125;(b)}
   \lbl{150,125;\bf{FSG3'x3'}}
   \lbl[W]{90,2;${}^{}$\hspace{15mm}}
   \lbl[W]{5,75,90;${}^{}$\hspace{15mm}}
   \end{lpic}
\end{minipage}
\begin{minipage}[c]{0.5\linewidth}
   \centering
   \begin{lpic}{Figures/ProdLongProfileRG(65mm)}
   \lbl{8,125;(c)}
   \lbl{150,125;\bf{RG115}}
   \lbl[W]{100,2;$x/x_{\mathrm{peak}}$}   
   \lbl[W]{2,75,90;$\mathcal{P}/\varepsilon_c^{\mathrm{iso}}$}
   \end{lpic}
\end{minipage}%
\begin{minipage}[c]{0.5\linewidth}
   \centering 
   \begin{lpic}{Figures/ProdLongProfileFG(65mm)}
   \lbl{8,125;(d)}
   \lbl{150,125;\bf{FSG3'x3'}}
   \lbl[W]{100,2;$x/x_{\mathrm{peak}}$}   
   \lbl[W]{5,75,90;${}^{}$\hspace{15mm}}
   \end{lpic}
\end{minipage}%
\caption{Longitudinal profiles of turbulent (a, b) transport and (c, d) production for three spanwise locations in the lee of the RG115, namely  (\protect\raisebox{-0.5ex}{\FilledSmallSquare}) $y=0$,  
(\ding{72}) $y=M/4$ and 
(\protect\raisebox{-0.5ex}{\FilledSmallTriangleRight}) $y=M/2$ (based on datasets 2 and 6, see table \ref{table:summary}). 
Additional data for RG115 from the $2 \times$XW experiments (datasets 9 and 10, see table \ref{table:summary}) are added to (a, c) and represented with open symbols.
Error bars  are described in the caption of figure \ref{fig:Um}. Dashed lines are polynomial fits to the data.}
\label{fig:TKELongProf}
\end{figure}

We now turn to the estimates of the main terms of the single-point turbulent kinetic energy (T.K.E.) transport equation,
\begin{align}
\underbrace{\frac{U_{k}}{2}\frac{\partial\, \overline{q^{2}}}{\partial x_{k}}}_{\mathcal{A}} = 
\underbrace{-\overline{u_{i} u_{j}}\, \frac{\partial U_{i}}{\partial x_{j}}}_{\mathcal{P}} 
\underbrace{-\frac{\partial}{\partial x_{k}}\left( \frac{\overline{u_{k} q^{2}}}{2} + \frac{\overline{u_{k} p}}{\rho} \right)}_{\mathcal{T}} +
\underbrace{\frac{\nu}{2} \frac{\partial^{2} \overline{q^{2}}}{\partial x^2_{m}}}_{\mathcal{D}_{\nu}} -
\,\varepsilon.
\label{eq:TKEa}
\end{align}
We have confirmed against all the data presented here that the Reynolds number is indeed sufficiently high ($Re_M=\mathcal{O}(10^5)$, $600<Re_{L^{1(1)}}<4000$) for the viscous diffusion $\mathcal{D}_{\nu}$ to be negligible relative to the turbulent dissipation, $\varepsilon$, similar to what was found in \cite{VV2011}.
We therefore now focus our attention on the transport and production terms $\mathcal{T}$ and $\mathcal{P}$.
The turbulent kinetic energy dissipation, $\varepsilon$, is estimated using the isotropic surrogate evaluated at the centreline, $\varepsilon_c^{\mathrm{iso}}$.
This choice is motivated in \S \ref{sec:Eps}.

The turbulent transport and production terms are estimated in a cylindrical coordinate system ($r,\phi,x$), where the $x$-coordinate is the same as the $x$-coordinate of our Cartesian coordinate system (see \S \ref{sec:apparatus}).
This is done with the aim to rely on an assumption of axisymmetry of the turbulent flow with respect to the centreline axis.
This assumption has recently been given substantial support for the decay region in the lee of FSGs by the wind tunnel measurements of \cite{Nagata2012}.
The turbulent production and triple velocity-correlation transport therefore take the form,
\begin{equation}
\mathcal{T} = - \frac{1}{r} \frac{\partial}{\partial r} \left(r\left(\frac{\overline{u_r q^2}}{2} + \frac{\overline{u_r p}}{\rho}\right) \right) - \frac{\partial}{\partial x} \left(\frac{\overline{u_1 q^2}}{2} + \frac{\overline{u_1 p}}{\rho}\right) 
\label{eq:T}
\end{equation}
and 
\begin{equation}
\mathcal{P} = - \overline{u_i u_1} \frac{\partial U_i}{\partial x} - \overline{u_i u_r} \frac{\partial U_i}{\partial r},
\label{eq:P}
\end{equation}
where $u_r$ and $u_1$ are the turbulent fluctuating velocity components aligned with $r$ and $x$, respectively.
Our axisymmetry assumption implies that we can estimate \eqref{eq:T} and \eqref{eq:P} by replacing $r$, $u_r$ and $\partial/ \partial r$ with $y$, $v\equiv u_2$ and $\partial/ \partial y$ in these equations.
Our experimental apparatus does not allow the measurements of the pressure-velocity correlations and thus only triple velocity-correlation turbulent transport is estimated.
Hence, what we are really calculating is:
\begin{equation}
\mathcal{T}= -{1\over y} {\partial \over \partial y} \left(y \frac{\overline{vq^{2}}}{2}\right) - {\partial \over \partial x} \left({\overline{uq^{2}}\over 2}\right) 
\label{eq:T2}
\end{equation}
where $q^{2} = u^{2} + 2v^{2}$ as in \S \ref{sec:2nd} and
\begin{equation}
\mathcal{P}= -\overline{u^{2}}{\partial U \over \partial x}  - \overline{uv}
{\partial U\over \partial y},
\end{equation}
since our flows are approximately parallel in the assessed region (i.e. $W\approx V\approx 0$).

Our conclusions in this subsection do not crucially depend on how good an approximation of the triple velocity-correlation transport and the turbulent production these two equations are. 
The quality of these approximations depends on how good the axisymmetry assumption is and on the impact of our neglect of the pressure-velocity correlation. 
This is a separate issue which we do not address here. 
What we do address here are the differences between the flows generated by RGs and FSGs and the effects of the
tunnel walls.
For simplicity, we refer to $\mathcal{T}$ defined by equation \eqref{eq:T2} as turbulent transport henceforth.

The spanwise profiles of turbulent transport normalised by $\varepsilon_{c}^{\mathrm{iso}}$ for the FSG3'x3'- and RG115-generated turbulence are shown in figure \ref{fig:TKESpanProf}a ($\varepsilon_{c}^{\mathrm{iso}}$ is $\varepsilon^{\mathrm{iso}}$ on the centreline).
The profiles in the lee of the two grids are recorded at four comparable downstream locations relative to $x_{\mathrm{peak}}$.
The prominent differences for the two turbulent flows are striking. 
Whereas the transport near the centreline for the FSG3'x3' is always negative (i.e. net loss of T.K.E.) and amounts to roughly 30\% of the dissipation throughout the assessed region of decay, for the RG115 at  the measurement station closest to $x_{\mathrm{peak}}$ it amounts to $\approx 45\%$ (i.e. net gain of T.K.E.),  changes sign further downstream and at the farthest measurement station becomes  relatively small ($> -10\%$; see also figure  \ref{fig:TKELongProf}a).
These differences are very likely caused by the geometrical differences between the grids. 
Note, however, that $Re_M$ for the FSG3'x3' recordings is about twice as those of the RG115 (see table \ref{table:summary}).
Nevertheless, at these Reynolds numbers, the variation in $Re_M$ cannot by itself justify the observed differences.  

Conversely, the differences in the spanwise profiles of turbulence production are more subtle (figure \ref{fig:TKESpanProf}b).
This observation is in-line with the more tenuous differences found in the spanwise profiles of $U$ and $\overline{uv}/u'v'$ (figures \ref{fig:Um}a,b and \ref{fig:ReyStress}c). 

The longitudinal profiles of the turbulence transport and production, at the centreline and at two parallel lines at $y=M/4,\,M/2$, can be found in figures \ref{fig:TKELongProf}a, b and show significant differences between the FSG and RG on the centreline, though less so off-centreline.
Complementary data from the $2\times$XW experiments (recorded at different downstream locations, c.f. table \ref{table:summary}) are also included for the RG115 case.
Note that these data are recorded at a lower inlet velocity, $U_{\infty}=10\mathrm{ms}^{-1}$ and consequently at a lower $Re_M$, but seem to follow roughly the same longitudinal profiles.
However, the difference in $Re_M$ (the $Re_M$ of the $2\times$XW data is about $2/3$ of the $Re_M$ of the single XW data) is insufficient to draw definitive conclusions concerning the Reynolds number dependence of the distribution and magnitude of the turbulent transport and production.

In \S \ref{sec:confinement} the effect of bounding wall confinement on the spanwise profiles of $U$ and $\overline{q^2}$ was demonstrated. 
Figures \ref{fig:TKESpanProf}c,d can be compared to figures \ref{fig:TKESpanProf}a,b to assess the confinement effects in terms of turbulent transport and production. 
It is clear that the effect of confinement is more pronounced on the turbulent transport profiles. 
For the RGs this leads to a change in the direction of the transport at the centreline close to the grid ($x/x_{\mathrm{peak}}\approx 1.4$) from $\mathcal{T}/\varepsilon_{c}^{\mathrm{iso}}\approx 45\%$ for RG115 to $\mathcal{T}/\varepsilon_{c}^{\mathrm{iso}}\approx -10\%$ for RG230 and an increase in the transport at $x/x_{\mathrm{peak}}\approx 1.8$ from $\mathcal{T}/\varepsilon_{c}^{\mathrm{iso}}\approx -5\%$ for RG115 to $\mathcal{T}/\varepsilon_{c}^{\mathrm{iso}}\approx -40\%$ for RG230.
For the FSGs the confinement leads to an improved `collapse' of the profiles with a value at the centreline of about $\mathcal{T}/\varepsilon_{c}^{\mathrm{iso}}\approx -45\%$ for the FSG18''x18'', c.a. $15\%$ higher than in FSG3'x3'.
The fact that this effect is felt throughout the decay leads to the hypothesis that it is caused by the influence of the wake/confining-wall interaction on the wake/wake interaction. 

On the other hand, the effect on the turbulent production is generally less pronounced, except far downstream for the FSG18''x18'' case where the profiles are severely distorted by comparison to the FSG3'x3' case, which can be attributed to the distortion in the mean velocity profiles (figure \ref{fig:ReyStressB}a). 
This is likely a consequence of the developing boundary layers on the confining walls as discussed in \S \ref{sec:confinement}. 
(Note that the last RG230 measurement is $x=3050$mm versus $x=4250$mm for FSG18''x18'', explaining why this effect is mostly seen for FSG18''x18''.)

Turbulent transport and production are shown to become small by comparison to the dissipation ($<10\%$) beyond $x/x_{\mathrm{peak}}=3.5$ for the RG115-generated turbulence, regardless of the spanwise location (see figures \ref{fig:TKELongProf}a,c).
A similar observation can be made for the FSG3'x3'-generated turbulence (see figures \ref{fig:TKELongProf}b,d), except for the turbulent transport around the centreline which has a substantially slower decay, perhaps only marginally faster than the dissipation and is persistent until the farthest downstream location measured.
A similar observation was previously presented in \cite{VV2011} for FSG18''x18''-generated turbulence.
With the present data we show that it is not due to confinement, although we also demonstrate that confinement does have a significant effect.

Finally, note that the presence of turbulent transport and production, particularly if varying with respect to the dissipation, influences the kinetic energy decay rate and its functional form. 
In \cite{VV2011} it was pointed out that the functional dependence on $x$ of the decaying turbulent kinetic energy is the same when the advection term in \eqref{eq:TKEa} is balanced by dissipation only and when it is balanced by dissipation and other terms provided that the ratio between dissipation and these other terms remains constant with $x$ during decay. 
This was indeed the case in the FSG18''x18'' centreline measurements of \cite{VV2011} where the one non-negligible term was the turbulent transport (including pressure transport in \citealt{VV2011}) which kept at an approximately constant fraction of
dissipation over the relevant streamwise length of measurements. 
This is not so, however, in the present centreline FSG3x3 measurements as can be seen in Figure \ref{fig:TKELongProf}b where the ratio of the turbulent transport to $\varepsilon_{c}^{\mathrm{iso}}$ varies significantly over the $x$-range of the reported measurements. 
These new data suggest that, if not caused by the pressure transport term, the constancy of turbulent transport as a fraction of dissipation may have been, at least partly, due to wall-confinement in the FSG18''x18'' case of \cite{VV2011}. 
The nature of the present FSG3'x3' set up is indeed one where wall confinement can be expected to be less of an issue.
 This would suggest that the power law fits reported in \cite{VV2011} for the FSG18''x18'' case of decaying turbulence ought to be different from those of decaying turbulence originating from our FSG3'x3' grid. 
 Our centreline data in figure \ref{fig:UmvsX} confirm this expectation as they are well fitted by $\overline{u^{2}} \sim (x-x_{0})^{-n}$ with $n=3.0$ and $x_{0}/x_{peak}= -1.4$ with the same nonlinear least-squares regression method used by \cite{VV2011} on their FSG18''x18'' data to obtain $n=2.41$ and $x_{0}/x_{peak}= -0.5$.
For reference, this same regression method yields $n=2.4$ and $x_{0}/x_{peak}= -1.1$ when applied to our RG115 data (see figure \ref{fig:UmvsX}).
(Note that \cite{KD2012} find $n\approx2.2$ for their regular grid.)

\section{Two-point large scale anisotropy} \label{sec:Lu}
We now turn our attention to the study of the large scale anisotropy of one of our decaying turbulent flows. 
We focus on RG115-generated turbulence for three reasons: (i) it is a better approximation to a periodic flow than
RG230-generated turbulence; (ii) it has a larger $x_{\mathrm{peak}}$ value and therefore a longer nonequilibrium decay region than RG60 in the wind tunnel; (iii) the constancy of the integral-length scale to the Taylor microscale ratio, which is indicative of the nonequilibrium region, is improved by comparison to FSG-generated turbulence as we report in \S \ref{sec:Llambda}.
Data from the farthest downstream location on the centreline of our RG60 set up is also shown. 
These data are from the equilibrium turbulence decay region because $x/x_{\mathrm{peak}} \approx 21 \gg 5$ \cite[see][]{VV2012}. As $x/x_{\mathrm{peak}}\approx 21$ corresponds to $x/M \approx 51$ for RG60, these data are also as close to homogeneous and isotropic turbulence as any of our data sets can be expected to get from knowledge of past measurements.

We study the downstream evolution of longitudinal and transverse correlations over both longitudinal and transverse
separations.
These correlations are defined as (no summation implied over the indices), 
\begin{equation}
B_{ii}^{(k)}(\mathbf{X},r) \equiv B_{ii}(\mathbf{X},r_k) = \frac{\overline{u_i(\mathbf{X}-r_k/2)\, u_i(\mathbf{X}+r_k/2)}}{\overline{u_i(\mathbf{X})\,u_i(\mathbf{X})}},
\end{equation}
where $r_k$ is the separation $r$  in the direction along the $x_k$-axis (with $x_1 = x$, $x_2 = y$, $x_3 = z$) and ${\mathbf{X}}$ is the position vector.

We use the data obtained with the $2\times$XW apparatus described in \S \ref{sec:apparatus} (datasets 14, 15 and 16 in table \ref{table:summary}) to calculate $B_{11}^{(2)}$ and $B_{22}^{(2)}$ and the time-varying signals to calculate $B_{11}^{(1)}$ and $B_{22}^{(1)}$ using Taylor's hypothesis.
We repeat these calculations for six downstream positions along the centreline which cuts through the centre of the
central mesh (see figure \ref{fig:grids}d) and six downstream positions along the line $(y=-M/2, z=0)$ which cuts through the lower bar of the central mesh. 

Though of less importance for this paper, it is nevertheless worth noting that these data can also be used to compute the scalar correlation function (with summation over $i$) $B_{ii}(\mathbf{X};r_{1},r_{2},0)$ in the $(r_{1}, r_{2}, 0)$ plane if the assumption is made that $B_{ii}(\mathbf{X}; r_{1},r_{2},0) = B_{\parallel} (\mathbf{X};r_{1},r_{2},0)+2B_{\perp\perp} (\mathbf{X};r_{1},r_{2},0)$, where $B_{\parallel}$ and $B_{\perp\perp}$ are the correlation functions of the velocity components parallel and perpendicular to $\mathbf{r}=(r_{1},r_{2},0)$.
By further assuming axisymmetry around the axis intercepting $\mathbf{X}$ and normal to the $(0,r_{2},r_{3})$ plane we
can then map $B_{ii} (\mathbf{X}; r_{1},r_{2},0 )$ onto the spherical coordinates $(R,\theta, \phi)$ (where $\phi$ is the angle around this axis) and extract an estimate of the spherically averaged correlation function,
\begin{equation}
B^*(\mathbf{X},r)\equiv \iint\limits_{r=|\mathbf{r}|}\!B_{ii}(\mathbf{X},\mathbf{r})\,d\mathbf{r}.
\end{equation}
The assessment of the two assumptions that we use to calculate $B^*$ lies beyond the scope of the present work as it concerns issues which are mostly peripheral to our main conclusions. 
Some support for these assumptions around the centreline can nevertheless be found in \cite{Sylvain2011} and \cite{Nagata2012}, though their validity around the $(y=-M/2, z=0)$ axis can be expected to be more doubtful. 
This section's main conclusions concern comparisons between the different longitudinal and transverse correlation functions and their associated integral-length scales $L_{ii}^{(k)}$ (no summation over $i$)
\begin{equation}
L_{ii}^{(k)}(\mathbf{X}) =\frac{1}{B_{ii}^{(k)}(\mathbf{X},0)}\int\limits_0^{\infty}\!B_{ii}^{(k)}(\mathbf{X},r)\,dr.
\end{equation}
We do also calculate the integral-length scale
\begin{equation}
L(\mathbf{X})= {1\over B^{*}(\mathbf{X},0)}\int_{0}^{\infty} B^{*}(\mathbf{X},r) dr
\end{equation}
and compare it with the other integral-length scales to check, for example, whether $3/2 L = 2L_{22}^{(1)}$ and/or $3/2 L=L_{11}^{(1)}$ as is the case in incompressible isotropic turbulence \cite[][and \citealp{Batchelor:book}]{MY75}. Our main checks, however, are to determine how far or close we are from the incompressible isotropic relations $2L_{22}^{(1)}=L_{11}^{(1)}$, $2L_{11}^{(2)}=L_{22}^{(2)}$, $B_{11}^{(1)}= B_{22}^{(2)}$ and $B_{11}^{(2)}= B_{22}^{(1)}$  \cite[]{MY75,Batchelor:book}.

\begin{figure}
\centering
\includegraphics[width=80mm]{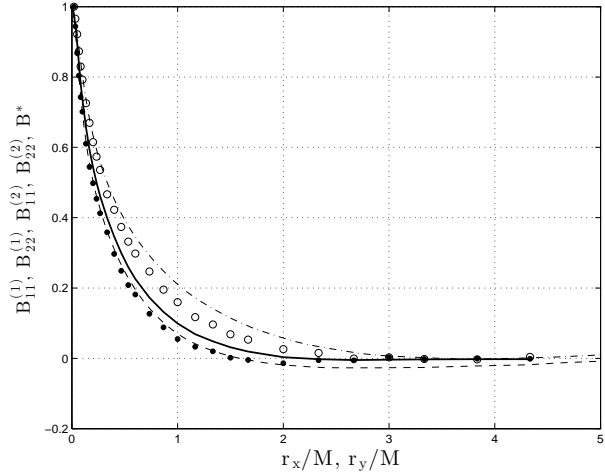}
\caption{Comparison between the longitudinal and transverse correlation functions for longitudinal and transverse separations of turbulence generated by RG60. The centroid of the correlation functions is located at  the centreline $y=0$ and at $x=3050\mathrm{mm}$, corresponding to $x/x_{\mathrm{peak}}\approx 21$. (dash-dotted line)  $\mathrm{B}_{11}^{(1)}$, (dashed line)  $\mathrm{B}_{22}^{(1)}$, (\protect\raisebox{-0.5ex}{\FilledSmallCircle}) $\mathrm{B}_{11}^{(2)}$,  (\protect\raisebox{-0.5ex}{\SmallCircle})  $\mathrm{B}_{22}^{(2)}$ and (solid line) $B^*$. $L_{11}^{(1)}/M = 0.67$, $L_{22}^{(2)}/L_{11}^{(1)} = 0.74$, $2\,L_{22}^{(1)}/L_{11}^{(1)} = 0.82$, $\,2L_{11}^{(2)}/L_{11}^{(1)} = 0.91$ and $3/2\,L/L_{11}^{(1)} = 0.87$. Dataset 11 (see table \ref{table:summary}) is used. }
\label{fig:B_RG60}
\end{figure}
\begin{figure}
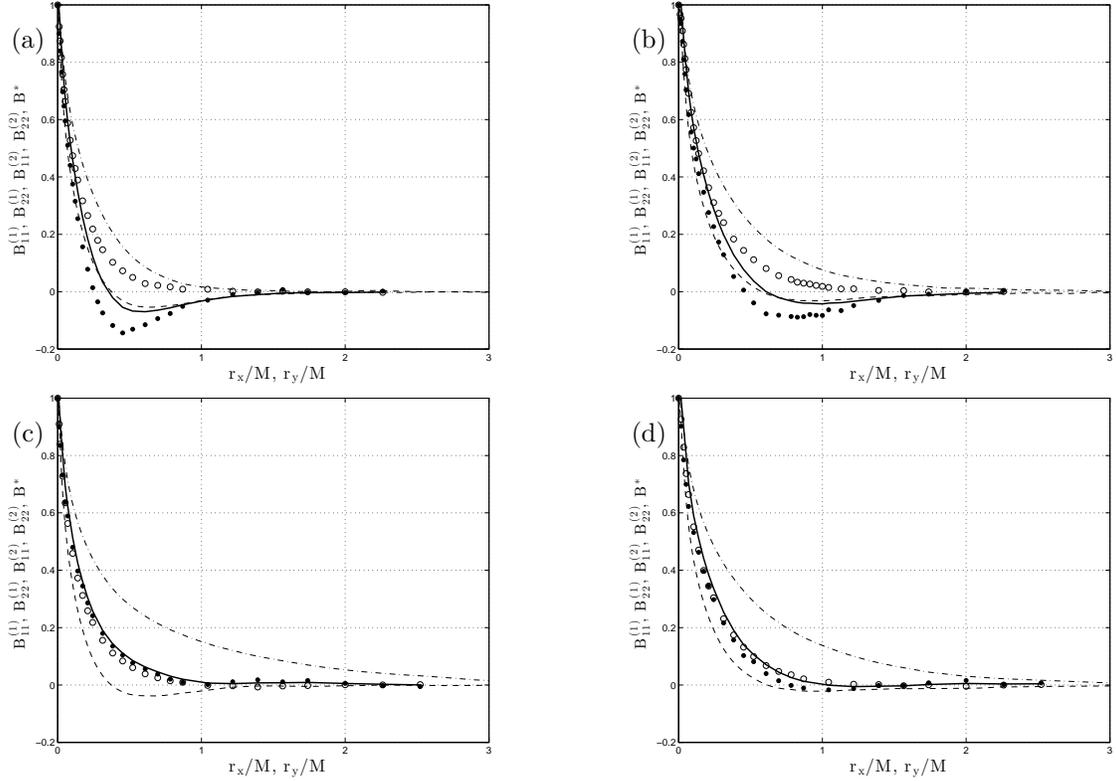

\centering
\begin{minipage}[c]{0.5\linewidth}
   \centering
   \begin{lpic}{Figures/CompareCorrCentre_x=1250(65mm)}
   \lbl{8,125;(a)}
   \end{lpic}
\end{minipage}%
\begin{minipage}[c]{0.5\linewidth}
   \centering 
   \begin{lpic}{Figures/CompareCorrCentre_x=3050(65mm)}
   \lbl{8,125;(b)}
   \end{lpic}
\end{minipage}
\begin{minipage}[c]{0.5\linewidth}
   \centering
   \begin{lpic}{Figures/CompareCorrOff_x=1250(65mm)}
   \lbl{8,125;(c)}
   \end{lpic}
\end{minipage}%
\begin{minipage}[c]{0.5\linewidth}
   \centering 
   \begin{lpic}{Figures/CompareCorrOff_x=3050(65mm)}
   \lbl{8,125;(d)}
   \end{lpic}
\end{minipage}%
\caption{Comparison between the longitudinal and transverse correlation functions for longitudinal and transverse separations of turbulence generated by RG115. The centroid of the correlation functions is located at, (a, b) the centreline $y=0$ and (c, d) behind a bar $y=-M/2$ for downstream locations, (a, c) $x=1250\mathrm{mm}$, corresponding to $x/x_{\mathrm{peak}}=1.5$ and (b, d) $x=3050\mathrm{mm}$, corresponding to $x/x_{\mathrm{peak}}=3.7$. Description of symbols/lines can be found in figure \ref{fig:B_RG60}.  Datasets 9 and 10 (see table \ref{table:summary}) are used. }
\label{fig:B_RG115}
\end{figure}


Figure \ref{fig:B_RG60} shows the different correlation functions for the RG60-generated turbulence at our farthest
downstream location on the centreline where the turbulence is expected to be closest to homogeneous and isotropic. 
For $r<2$M the transverse correlations are roughly equal, $B_{22}^{(1)} \approx B_{11}^{(2)}$, whereas the longitudinal correlations, $B_{11}^{(1)}$ and $B_{22}^{(2)}$ are less so. 
For $r>2$M, $B_{22}^{(1)}$ seems to tend slowly to zero, contrasting with $B_{11}^{(2)}$.
These departures between $B_{11}^{(2)}$ and  $B_{22}^{(1)}$ for large $r$ may be related to the lack of validity of Taylor's hypothesis for long time differences, but may also be genuine departures from isotropy.

The ratio between the different integral-length scales is presented in the caption of figure  \ref{fig:B_RG60} (for isotropic turbulence these ratios are equal to one) and they indicate small, but non-negligible, departures from isotropy even at this relatively far downstream location ($x/M\approx 51$).

The RG115 data, on the other hand, show a larger departure from isotropy (figure \ref{fig:B_RG115} and table \ref{table:Lu}). 
The ratios $L_{22}^{(2)}/L_{11}^{(1)}$ and $2L_{22}^{(1)}/L_{11}^{(1)}$ do not show any tendency towards isotropy between $x/x_{\mathrm{peak}} \approx 1.1$ and $x/x_{\mathrm{peak}} \approx 3.7$ on the centreline as they both remain about constant with values around $0.62$ and $0.54$ respectively. 
The ratio $2L_{11}^{(2)}/L_{11}^{(1)}$ is even further away from the isotropic value $1$ but grows quite steeply with streamwise distance $x$ on the centreline. 
The very small values of this particular ratio reflect the prominent negative loop in $B_{11}^{(2)}$ at the lower $x/x_{\mathrm{peak}}$ locations, a negative loop which progressively weakens as $x/x_{\mathrm{peak}}$ increases thereby yielding increasing values of $2L_{11}^{(2)}/L_{11}^{(1)}$. 
This effect also presumably explains the steep growth of $3/2 L/L_{11}^{(1)}$ with increasing $x/x_{\mathrm{peak}}$ along the centreline because of the related negative loop in $B^{*}$ at the lower $x/x_{\mathrm{peak}}$ locations which
also disappears with increasing $x/x_{\mathrm{peak}}$.

The decaying oscillation and related negative loop in the transverse correlation $B_{11}^{(2)}$ of longitudinal fluctuating turbulent velocities is likely a remnant of the periodicity of the grid leading to a peak in negative correlation mid way between bars, i.e. at $r=M/2$ as indeed observed in figure \ref{fig:B_RG115}. 
The grid's periodicity can indeed leave a mark on the flow in the form of a transverse near-periodicity of its vortex  shedding which disappears far downstream. 
When this happens and the correlation function has a negative loop as a result, the integral-length scale obtained by
integrating this correlation function loses its usual meaning as a spatial extent of correlation.

Along the $(y=-M/2, z=0)$ line crossing the lower bar of the central mesh of the grid the length-scale ratios are different. Firstly, $L_{22}^{(2)}/L_{11}^{(1)}$ and $2L_{22}^{(1)}/L_{11}^{(1)}$ exhibit a significant increase with increasing $x$ which they do not exhibit on the centreline. 
However, $L_{22}^{(2)}/L_{11}^{(1)}$ and $2L_{22}^{(1)}/L_{11}^{(1)}$ take significantly lower values than at the same streamwise positions along the centreline, indicating more anisotropy in the wake of the bar than along the centreline between bars. 
By a different measure, though, that of $2L_{11}^{(2)}/L_{11}^{(1)}$, the turbulence appears more isotropic in the wake of the bar than along the centreline because $2L_{11}^{(2)}/L_{11}^{(1)}$ is very much closer to 1 behind the bar. 
It is clear that each of the ratios between different length-scales and $L_{11}^{(1)}$ show a different trend as the turbulence decays.
Nevertheless, it is also clear that $B_{11}^{(1)} > B_{22}^{(2)}$ for all separations $r$, and consequently $L_{11}^{(1)} > L_{22}^{(2)}$ at all streamwise positions accessed by our measurements both behind the bar and along the centreline.
This suggests that the large-scale eddies are elongated in the streamwise direction. 
This had previously been observed for RG-generated turbulence by \cite{Graham} which estimated that $L_{33}^{(2)}/L_{33}^{(1)}\approx 3/4$ at $x/M\approx 11$, i.e. a $25\%$ elongation in the streamwise direction.
Also note that the PIV data of \cite{discettietal11}, taken along the centreline of a FSG similar to our FSG18''x18'' also suggest such an elongation of large-scale eddies in the streamwise direction (see their figure 14).

\begin{table}
\centering
\begin{tabular*}{0.9\textwidth}{@{\extracolsep{\fill}}cccccccc}
\vspace{1mm}$x/x_{\mathrm{peak}}$ &   & 1.1 & 1.5 & 2.0 & 2.6 & 3.1 & 3.7 \\
\hline
\vspace{.5mm}   \multirow{6}{*}{\begin{sideways}Centreline\end{sideways}} & $L_{11}^{(1)}/M$& 0.24 & 0.25 & 0.27 & 0.30 & 0.34 & 0.36 \\ 
\vspace{.5mm}                   & $L_{22}^{(2)}/L_{11}^{(1)}$     & 0.60 & 0.64 & 0.65 & 0.64 & 0.61 & 0.62 \\
\vspace{.5mm}                   & $2\,L_{22}^{(1)}/L_{11}^{(1)}$& 0.54 & 0.53 & 0.54 & 0.54 & 0.53 & 0.55 \\
\vspace{.5mm}                   & $2\,L_{11}^{(2)}/L_{11}^{(1)}$& 0.13 & 0.14 & 0.19 & 0.30 & 0.36 & 0.44 \\
\vspace{2mm}                    & $3/2\,L/L_{11}^{(1)}$              & 0.40 & 0.44 & 0.48 & 0.54 & 0.57 & 0.61 \\
\vspace{0.5mm}   \multirow{6}{*}{\begin{sideways}Behind bar\end{sideways}}& $L_{11}^{(1)}/M$ & 0.40 & 0.40 & 0.42 & 0.40 & 0.43 & 0.44 \\
\vspace{.5mm}                  & $L_{22}^{(2)}/L_{11}^{(1)}$    & 0.38 & 0.40 & 0.40 & 0.46 & 0.44 & 0.47 \\
\vspace{.5mm}                  & $2\,L_{22}^{(1)}/L_{11}^{(1)}$& 0.31 & 0.34 & 0.36 & 0.43 & 0.44 & 0.49 \\
\vspace{.5mm}                  & $2\,L_{11}^{(2)}/L_{11}^{(1)}$& 0.95 & 0.94 & 0.88 & 0.91 & 0.86 & 0.84 \\
\vspace{.5mm}                  & $3/2\,L/L_{11}^{(1)}$              & 0.75 & 0.75 & 0.74 & 0.79 & 0.74 & 0.75 \\
\end{tabular*}
\caption{Several integral length-scales for different downstream location for the RG115-generated turbulence. The different integral length-scales are normalised with $L_{11}^{(1)}$ with a pre-factor such that unity would correspond to isotropic incompressible turbulence. Datasets 9 and 10 (see table \ref{table:summary}) are used.}
\label{table:Lu}
\end{table}

\section{Small-scale anisotropy and dissipation rate estimates} \label{sec:Eps}

We now turn the focus to the (an)isotropy of the small scales.
These are customarily assessed by comparing the ratios between the various mean square velocity derivatives with the isotropic benchmark \cite[see e.g.][]{George91}.
For example,
\begin{equation}
K_1 = 2\frac{\overline{\left(\partial u/\partial x\right)^2}}{\overline{\left(\partial v/\partial x\right)^2}}, \hspace{5mm} K_2 = 2\frac{\overline{\left(\partial v/\partial y\right)^2}}{\overline{\left(\partial u/\partial y\right)^2}}, \hspace{5mm}
K_3 = 2\frac{\overline{\left(\partial u/\partial x\right)^2}}{\overline{\left(\partial u/\partial y\right)^2}},
\label{eq:localiso}
\end{equation}
should all be unity for a locally isotropic flow \cite[see][where all the velocity derivative ratios are derived for an isotropic turbulent field]{Taylor1935}. 

There have been many experimental investigations of local isotropy in canonical turbulent shear flows \cite[see e.g.][and references therein]{BAS1987,George91,SV94}, but it seems that in this context grid-generated turbulence has not attracted much attention, perhaps because local isotropy is thought to be guaranteed.  
However, the experimental data by \cite{Tsinober92} for RG-generated turbulence suggest significant departures from local isotropy (to the best of our knowledge no other assessment of local isotropy for RGs can be found in the literature).  
These data have significant scatter making it difficult to discern trends as the flow decays, particularly for $K_2$, but it seems that $K_1$ and $K_3$ are about constant between $6<x/x_{\mathrm{peak}}<31$ with numerical values surrounding $1.4$ and $0.8$ respectively. 
Recently, \cite{gomesfernandesetal12} presented estimates of $K_1$ and $K_3$ in the lee of three FSGs along the centreline up to a downstream region of about $4x_{\mathrm{peak}}$.
Their data for the FSG similar to the present FSG18''x18'' indicate that $K_1$ and $K_3$ are approximately constant beyond $x_{\mathrm{peak}}$ with  numerical values of about $1.2$ and $1.1$, respectively.

The validity of the approximations of local isotropy (or local axisymmetry) in the decaying turbulence generated by square-mesh grids is, however, a peripheral topic to the present work. 
What is, in fact, the main concern here is to assess how the anisotropy of the small scales varies as the turbulent flow decays and/or $Re_{\lambda}$ changes. 
If the ratios $K_1$, $K_2$ and $K_3$ (and/or ratios than can be formed with the other components of the mean square velocity derivative tensor) vary significantly during the turbulence decay and/or with $Re_{\lambda}$, then the surrogate isotropic dissipation estimate, $\varepsilon^{\mathrm{iso}} = 15\nu \overline{(\partial u/\partial x)^2}$, obtained from 
one-component measurements (e.g. with a single hot-wire)  is not representative of the true turbulent kinetic energy dissipation, $\varepsilon$, as usually assumed.
This would bear severe consequences, not only for the present work, but also for turbulence research in general since the overwhelming majority of the dissipation estimates found in the literature are indeed estimates of the surrogate $\varepsilon^{\mathrm{iso}}$, typically obtained with a single hot-wire. Multicomponent hot-wires and particle image velocimetry, PIV, have resolution and/or noise issues and inhibit accurate and reliable measurements of several components of the mean square velocity derivative tensor \cite[see also the discussion in][where an exact filter is proposed for the unavoidable noise contaminating PIV measurements]{Anthony2012}.
In any case, the data of \cite{Tsinober92} and \cite{gomesfernandesetal12} do not suggest that there are significant variations of the anisotropy ratios.  
   
We use the $2\times$XW apparatus described in \S \ref{sec:apparatus} (datasets 14 \& 16 in table \ref{table:summary}) to measure $\overline{\left(\partial u/\partial x\right)^2}$, $\overline{\left(\partial v/\partial x\right)^2}$, $\overline{\left(\partial u/\partial y\right)^2}$ and $\overline{\left(\partial v/\partial y\right)^2}$ for the RG60- and RG115-generated turbulent flows along the centreline, which in turn allow the estimation of the ratios \eqref{eq:localiso}.
These are presented in tables \ref{tab:RG60} and \ref{tab:RG115}, respectively. 
The first observation is that the ratios $K_1$  and $K_3$ are roughly constant during the turbulence decay for both the  RG60 and the RG115 data, in-line with the observations from the data of \cite{gomesfernandesetal12}. 
The numerical values of $K_1\approx 1.09\,\&\,1.04$ and $K_3\approx 0.8\,\&\,0.72$ for the RG60 and RG115 data respectively suggest that the RG115 data are closer to the isotropic benchmark in terms of the $K_1$ ratio but conversely, the RG60 data are closer to unity in terms of the $K_3$ ratio. 
The ratio $K_2$ increases away from unity, particularly for the RG60 data, which could be an indication of increasing anisotropy as the flow decays. However, $K_2$ is a ratio involving $\overline{(\partial v/\partial y)^2}$ whose measurement is strongly contaminated by aerodynamic interference (see \S \ref{sec:resolution}) and therefore the results are likely to be artificial.

Comparing with the data of \cite{gomesfernandesetal12}, it is clear that the present  numerical values of $K_1$ are always closer to the isotropic benchmark. 
Curiously, the ratio $K_3$ for the present data are $20\%$ to $30\%$ smaller than 1 whereas for the data of \cite{gomesfernandesetal12} they are about $10\%$ higher than unity. 
These differences may be attributable to the different inflow conditions, e.g. grid geometry, free-stream turbulence and $Re_M$, but may also be an artifice of measurement bias.  
In any case, the present data and those of \cite{gomesfernandesetal12} support the hypothesis that the small-scale anisotropy remains approximately constant.

The present data allow the calculation of several estimates of the turbulent dissipation.
In particular, the four mean square velocity derivative components are necessary and sufficient to estimate the dissipation in a locally axisymmetric turbulent flow \cite[]{George91}.
Even though the present data do not allow the test for local axisymmetry, one might nevertheless expect a locally axisymmetric dissipation estimate to be closer to the actual dissipation rate than the isotropic dissipation estimate.
%
\begin{table}
\centering
\begin{tabular*}{0.9\textwidth}{@{\extracolsep{\fill}}rcccccc}
Location & $1250$ & $1700$ & $2150$ & $2600$ & $3050$ \\
\hline
$x/x_{\mathrm{peak}}$ & $8.5$ & $11.5$ & $15.6$ & $17.6$ & $20.7$ \\
$Re_{\lambda}^{\mathrm{iso}}$ & $106$ & $102$ & $98$ & $94$ & $91$ \\
$Re_{\lambda}$ & $91$ & $86$ & $82$ & $80$ & $77$ \\
$\overline{q^2}\,(\mathrm{m^2s^{-2}})$  & $0.43$ & $0.28$ & $0.20$ & $0.16$ & $0.13$ \\
$\lambda \, (\mathrm{mm})$ & $3.6$ & $4.2$ & $4.8$ & $5.3$ & $5.6$ \\
$\eta\, (\mathrm{mm})$ & $0.19$ & $0.23$ & $0.27$ & $0.30$ & $0.32$ \\
$\varepsilon^{\mathrm{iso}}$   & 2.19 &  0.99 & 0.55 & 0.36 & 0.25 \\
$\varepsilon^{\mathrm{iso},2}$& 2.05 &  0.92 & 0.51 & 0.33 & 0.24 \\
$\varepsilon^{\mathrm{iso},3}$& $2.51\pm0.07$ &  $1.18\pm0.02$ & $0.66\pm0.02$ & $0.42\pm0.01$ & $0.30\pm0.01$ \\
$\varepsilon^{\mathrm{axi}}$  & $2.75\pm 0.19$ &  $1.32\pm0.08$ & $0.77\pm0.07$ & $0.51\pm0.05$ & $0.36\pm0.04$  \\
$\Delta/\eta \approx l_w/\eta$ & 2.6  & 2.2 & 1.9 & 1.7 & 1.6 \\
$(\Delta y\approx 1.2\mathrm{mm})/\eta$& 6.4 &  5.4   &  4.8  & 4.0   & 3.8 \\
$(\Delta y\approx 2.0\mathrm{mm})/\eta$& 10.5 &  9.0   &  7.5  & 6.7   & 6.2 \\
$K_1$& 1.09 & 1.09 & 1.08 & 1.09 & 1.09 \\
$K_2$& $1.18\pm 0.13$ &  $1.21\pm 0.13$ & $1.31\pm 0.18$ & $1.46\pm 0.22$ & $1.41\pm 0.23$ \\
$K_3$& $0.83\pm 0.03$ &  $0.79\pm 0.02$ & $0.79\pm 0.01$ & $0.83\pm 0.03$ & $0.81\pm 0.01$ \\
$u'/v'$ & $1.07$ & $1.07$ & $1.06$ & $1.06$ & $1.06$ \\
\end{tabular*} 
\caption{Turbulence statistics for the RG60 along the centreline. The dissipation estimate $\varepsilon^{\mathrm{iso},3}$ is used to compute $Re_{\lambda}$, $\lambda$ and $\eta$, whereas $\varepsilon^{\mathrm{iso}}$ is used to compute $Re_{\lambda}^{\mathrm{iso}}$. Dataset 16 (see table \ref{table:summary}) is used.}
\label{tab:RG60}
\end{table} 
%
\begin{table}
\centering
\begin{tabular*}{0.9\textwidth}{@{\extracolsep{\fill}}rcccccc}
Location & $1250$ & $1700$ & $2150$ & $2600$ & $3050$ \\
\hline
$x/x_{\mathrm{peak}}$ & $1.5$ & $2.0$ & $2.6$ & $3.1$ & $3.7$ \\
$Re_{\lambda}^{\mathrm{iso}}$  & $156$ & $140$ & $133$ & $124$ & $116$ \\
$Re_{\lambda}$  & $114$ & $105$ & $98$ & $91$ & $88$ \\
$\overline{q^2}\,(\mathrm{m^2s^{-2}})$  & $0.78$ & $0.51$ & $0.36$ & $0.26$ & $0.20$ \\
$\lambda \, (\mathrm{mm})$ & $3.3$ & $3.8$ & $4.3$ & $4.7$ & $5.2$ \\
$\eta\, (\mathrm{mm})$ & $0.16$ & $0.19$ & $0.22$ & $0.25$ & $0.28$ \\
$\varepsilon^{\mathrm{iso}}\,(\mathrm{m^{2}s^{-3}})$  &  4.23 & 2.06 & 1.17 & 0.70 & 0.45 \\
$\varepsilon^{\mathrm{iso},2}\,(\mathrm{m^{2}s^{-3}})$&  4.08 & 2.02 & 1.13 & 0.67 & 0.43 \\
$\varepsilon^{\mathrm{iso},3}\,(\mathrm{m^{2}s^{-3}})$&  $5.21\pm0.24$ & $2.59\pm0.12$ & $1.48\pm0.02$ & $0.89\pm0.02$ & $0.55\pm0.04$\\
$\varepsilon^{\mathrm{axi}}\,(\mathrm{m^{2}s^{-3}})$  &  $5.44\pm0.58$ & $2.71\pm0.21$ & $1.59\pm0.07$ & $0.97\pm0.06$ & $0.62\pm0.09$  \\
$\Delta/\eta \approx l_w/\eta$ & 3.2 & 2.6 & 2.3 & 2.0 & 1.8 \\
$(\Delta y\approx 1.2\mathrm{mm})/\eta$& 7.7 & 6.3 & 5.5 & 5.7 & 4.2 \\
$(\Delta y\approx 2.0\mathrm{mm})/\eta$& 12.5& 10.8 & 9.3 & 8.1 & 7.2 \\
$K_1$ & 1.05 & 1.03 & 1.04 & 1.05 & 1.04 \\
$K_2$ &  $1.01\pm0.05$ & $1.00\pm0.13$ & $1.06\pm0.08$ & $1.09\pm0.11$ & $1.20\pm0.21$\\
$K_3$ &  $0.73\pm0.01$ & $0.72\pm0.04$ & $0.72\pm0.01$ & $0.71\pm0.01$ & $0.77\pm0.02$\\
$u'/v'$ & $1.18$ & $1.15$ & $1.16$ & $1.16$ & $1.14$ \\
\end{tabular*} 
\caption{Turbulence statistics for the RG115  along the centreline. The dissipation estimate $\varepsilon^{\mathrm{iso},3}$ is used to compute $Re_{\lambda}$, $\lambda$ and $\eta$, whereas $\varepsilon^{\mathrm{iso}}$ is used to compute $Re_{\lambda}^{\mathrm{iso}}$. Dataset 14 (see table \ref{table:summary}) is used.}
\label{tab:RG115}
\end{table}

The data are used to calculate four estimates of the dissipation, namely 
\begin{equation}
\left\lbrace
\begin{aligned}
\varepsilon^{\mathrm{iso}} &\equiv 15\nu \overline{(\partial u/\partial x)^2};\\
 \varepsilon^{\mathrm{iso,2}} &\equiv \nu(3\overline{(\partial u/\partial x)^2}+6\overline{(\partial v/\partial x)^2});\\
\varepsilon^{\mathrm{iso,3}} &\equiv \nu(\overline{(\partial u/\partial x)^2}+2\overline{(\partial v/\partial x)^2}+4\overline{(\partial u/\partial y)^2}+2\overline{(\partial v/\partial y)^2});\\
\varepsilon^{\mathrm{axi}} &\equiv \nu(-\overline{(\partial u/\partial x)^2}+2\overline{(\partial v/\partial x)^2}+2\overline{(\partial u/\partial y)^2}+8\overline{(\partial v/\partial y)^2}).
\end{aligned}
\right.
\end{equation}
The first estimate, $\varepsilon^{\mathrm{iso}}$ is the widely used isotropic dissipation estimate where all the kinematic constraints of locally isotropy are implied. 
The second estimate ($\varepsilon^{\mathrm{iso,2}}$) is very similar to the first, but with one less isotropy relation, namely $K_1 = 1$ is not used.
The last estimate ($\varepsilon^{\mathrm{axi}}$) is the locally axisymmetric estimate  \cite[]{George91}. 
However, this dissipation estimate heavily weights $\overline{(\partial v/\partial y)^2}$ whose measurement is, as noted before, strongly contaminated by aerodynamic interference (see \ref{sec:resolution}).
To overcome this limitation, a more robust estimate ($\varepsilon^{\mathrm{iso,3}}$) that uses all the measured velocity derivative components, but inevitably still relies on isotropy, is proposed.  The estimated $\varepsilon^{\mathrm{iso,3}}$ is a natural extension of $\varepsilon^{\mathrm{iso,2}}$ without assuming that $K_2=1$ nor that $\overline{\left(\partial u/\partial x\right)^2}=\overline{\left(\partial v/\partial y\right)^2}$.
Note that $\overline{\left(\partial u/\partial y\right)^2}$  and $\overline{\left(\partial v/\partial y\right)^2}$ can be computed from datasets with different spanwise separations between the X-probes (see \S \ref{sec:resolution}). 
It was seen in \S \ref{sec:resolution} that as the separation between the X-probes is increased the interference between the probes is reduced but the deterioration of the resolution causes a bias to the measured value which can partly be compensated using the correction factors proposed by \cite{ZA96}. However, if the correction factor is excessive one cannot expect the method to yield accurate results.  
For the present estimates of $\varepsilon^{\mathrm{iso,3}}$ and $\varepsilon^{\mathrm{axi}}$ it was deemed that the best compromise between interference and resolution for the present data was given by probe separations between $\Delta y=1.6$ and $\Delta y=2.5$. We compute the average value of the estimates obtained for the three datasets at $\Delta y=1.6,\, 2.0\,\&\, 2.5$ and present the difference as an uncertainty interval in tables \ref{tab:RG60} and \ref{tab:RG115}. Note that in that uncertainty interval the statistical convergence has not been included, but it is estimated to be less than $\pm 1\%$ (\emph{c.f.} \S \ref{sec:conv}).

Taking $\varepsilon^{\mathrm{iso,3}}$ as the benchmark, it is noticeable that throughout the turbulence decay, the isotropic dissipation estimate, $\varepsilon^{\mathrm{iso}}$, underestimates dissipation rate by about $20\%$ for the RG115 data and $15\%$ for the RG60 data, whereas the estimate $\varepsilon^{\mathrm{iso,2}}$, underestimates dissipation rate by c.a. $30\%$ for the RG115 data and $26\%$ for the RG60 data
This motivates the choice of $\varepsilon^{\mathrm{iso}}$ rather than $\varepsilon^{\mathrm{iso,2}}$ as the prime dissipation estimate whenever the additional data needed to estimate $\varepsilon^{\mathrm{iso,3}}$ are not available. 
Most importantly, the observation that $\varepsilon^{\mathrm{iso}}$, $\varepsilon^{\mathrm{iso,2}}$ and $\varepsilon^{\mathrm{iso,3}}$ (and $\varepsilon^{\mathrm{axi}}$ within the scatter) remain approximately proportional throughout the decay lead to the expectation that these are also approximately proportional to the actual dissipation rate.
Therefore, using either of the dissipation estimates to infer, for example, on the behaviour of the normalised energy dissipation rate (see \S \ref{sec:Llambda}) leads to curves with the same functional form but offset from one another.

\section{Energy dissipation scaling} \label{sec:Llambda}
\begin{figure}
\centering
\begin{minipage}[c]{1\linewidth}
   \centering
   \begin{lpic}{Figures/CepsLongL(120mm)}
   \lbl{8,125;(a)}
   \lbl[W]{5,75,90;\hspace{10mm}$2L_{22}^{(1)}/\lambda$, $L_{11}^{(1)}/\lambda$\hspace{10mm}}
   \lbl[W]{100,3;$Re_\lambda$}
   \end{lpic}
\end{minipage}
\begin{minipage}[c]{1\linewidth}
   \centering 
   \begin{lpic}{Figures/CepsTransL(120mm)}
   \lbl{8,125;(b)}
   \lbl[W]{5,75,90;\hspace{10mm}$2L_{11}^{(2)}/\lambda$, $L_{22}^{(2)}/\lambda$, $3L/2\lambda$\hspace{10mm}}
   \lbl[W]{100,3;$Re_\lambda$}
   \end{lpic}
\end{minipage}
\caption{Downstream evolution of the ratio between various integral-length scales and the Taylor microscale versus $Re_{\lambda}$ in the lee of RG115 for $1.1<x/x_{\mathrm{peak}}<3.7$ and $U_{\infty}=10\mathrm{ms}^{-1}$ (see table \ref{table:summary}). 
 The integral-length scales are based on the longitudinal and transverse correlations for (a) longitudinal and (b) transverse separations.
 (\protect\raisebox{-0.5ex}{\FilledSmallSquare} $\!|\!\!$ \protect\raisebox{-0.5ex}{\rlap{\SmallSquare}\SmallCross} $\!|\!\!$ \protect\raisebox{-0.5ex}{\SmallSquare}) $L_{11}^{(1)}/\lambda$ for $y/M=0,\,-0.25,\,-0.5$, (\protect\raisebox{-0.5ex}{\FilledSmallCircle} $\!|\!\!$ \protect\raisebox{-0.4ex}{\rlap{\Circle}\SmallCross} $\!|\!\!$ \protect\raisebox{-0.5ex}{\SmallCircle}) $L_{22}^{(1)}/\lambda$ for $y/M=0,\,-0.25,\,-0.5$, (\protect\raisebox{-0.5ex}{\FilledDiamondshape} $\!|\!\!$ \protect\raisebox{-0.5ex}{\Diamondshape}) $L_{11}^{(2)}/\lambda$ for $y/M=0,\,-0.5$, (\ding{72} $\!|\!\!$ \ding{73}) $L_{22}^{(2)}/\lambda$ for $y/M=0,\,-0.5$ and (\protect\raisebox{-0.5ex}{\FilledSmallTriangleUp} $\!|\!\!$ \protect\raisebox{-0.5ex}{\SmallTriangleUp}) $2L/\lambda$ for $y/M=0,\,-0.5$.
The dashed line follows $B/15\,Re_{\lambda}$, with (a) $B=1.4$ and (b) $B=1.0$. 
The dotted line in (a) follows $A + B/15\,Re_{\lambda}$  with $A=2.2$ and $B=1.0$.
$\varepsilon^{\mathrm{iso}}$ is used as a surrogate for the dissipation rate due to the lack of estimates of $\varepsilon^{\mathrm{iso,3}}$ for off-centreline locations. Datasets 9 and 10 (see table \ref{table:summary}) are used.}
\label{fig:Llambda}
\end{figure}

\begin{figure}
\centering
\includegraphics[width=120mm]{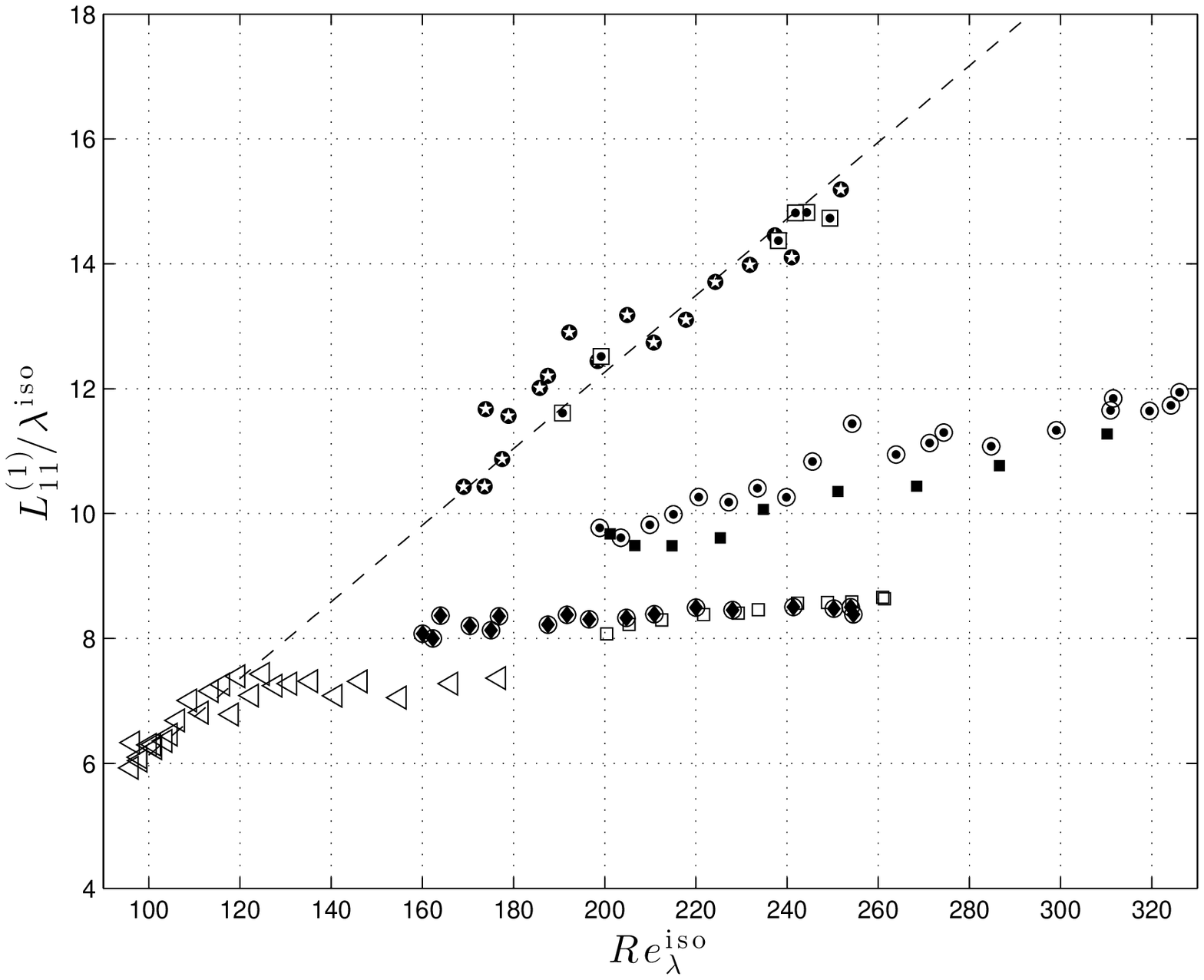}
\caption{Downstream evolution of $L_{11}^{(1)}/\lambda^{\mathrm{iso}}$ versus $Re_{\lambda}^{\mathrm{iso}}$ during turbulence decay. Data at the centreline ($y=0$):  
(\protect\raisebox{-0.5ex}{\SmallTriangleLeft}) RG60,
(\protect\raisebox{-0.5ex}{\rlap{\Circle}\FilledDiamondshape}) RG115, 
(\protect\raisebox{-0.5ex}{\SmallSquare}) RG230,
(\protect\raisebox{-0.5ex}{\rlap{\BigCircle}\FilledSmallCircle}) FSG3'x3' and
(\protect\raisebox{-0.5ex}{\FilledSmallSquare}) FSG18''x18''. 
Data behind the bar ($y=M/2$):
(\ding{74}) RG115 and
(\protect\raisebox{-0.5ex}{\rlap{\BigSquare}\FilledSmallCircle}) FSG3'x3'.
The dashed line follows $B/15\,Re_{\lambda}$, with $B=0.92$. Datasets 1, 4, 7, 8, 12 and 13 (see table \ref{table:summary}) are used.}
\label{fig:LlambdaB}
\end{figure}

Since \cite{Taylor1935}, the kinetic energy dissipation rate per unit mass $\varepsilon$ is scaled with the turbulent kinetic energy and length-scale of the large eddies. 
There is no ambiguity in this definition for isotropic turbulence where $3/2 L = 2L_{22}^{(1)} =L_{11}^{(1)}$ and therefore $\varepsilon = C_{\varepsilon} (\overline{q^{2}}/3)^{3/2}/L_{11}^{(1)}$ with a potential dependence of the dimensionless coefficient $C_{\varepsilon}$ on Reynolds numbers which does not change qualitatively if, in the definition of $C_{\varepsilon}$, $L_{11}^{(1)}$ is replaced by one of the other integral length-scales.

However, when the large-scale eddies are "elongated"/anisotropic and characterised by different integral scales $L_{11}^{(1)}$ and $L_{22}^{(2)}$ in different directions as found in the previous section, then the dependence of $C_{\varepsilon}$ on Reynolds numbers may depend on the choice of length-scale in its definition. 
One can indeed define the coefficients
\begin{equation}
C_{\varepsilon}^{i(k)} \equiv \varepsilon {L_{ii}^{(k)}\over (\overline{q^{2}}/3)^{3/2}}.
\end{equation}
where $L_{ii}^{(k)}/ (\overline{q^{2}}/3)^{1/2}$ are the various time-scales corresponding to the various integral length-scales. 
In this section we use our data to compare how some of these coefficients depend on Reynolds number.

The ratio between the integral length-scale and the Taylor microscale is directly related to the normalised energy dissipation rate.
From the general definitions of the Taylor microscale $\lambda$($\equiv (5 \nu \overline{q^2}/\varepsilon)^{1/2}$) and $Re_{\lambda}$($\equiv (\overline{q^2}/3)^{1/2}\lambda/\nu$) it follows that (no summation over $i$ is implied),
\begin{equation}
C_{\varepsilon}^{i(k)}= \frac{15}{Re_{\lambda}}\frac{L_{ii}^{(k)}}{\lambda}.
\label{eq:Ceps}
\end{equation}
The dimensionless dissipation constant $C_{\varepsilon}^{i(k)}$ is independent of local Reynolds number $Re_{\lambda}$ if and only if $L_{ii}^{(k)}/\lambda$ is proportional to $Re_{\lambda}$; it is proportional to $1/Re_{\lambda}$ if and only if $L_{ii}^{(k)}/\lambda$ remains constant during turbulence decay and $L_{ii}^{(k)}/\lambda$ is therefore independent of $Re_{\lambda}$.

Note that for off-centreline measurements only two mean-square velocity gradients are measured, $\overline{(\partial u/\partial x)^2}$ and $\overline{(\partial v/\partial x)^2}$. Based on the results from \S \ref{sec:Eps} we choose $\varepsilon^{\mathrm{iso}}$ as our dissipation estimate.  Since $\varepsilon/ \varepsilon^{\mathrm{iso}} \approx \mathrm{constant}$, at least for the centreline, the functional form of $L_{ii}^{(k)}/\lambda$ versus $Re_{\lambda}$ does not meaningfully change, and therefore the physical interpretation is not jeopardised.
Instead, the fact that we always underestimate $\varepsilon$ by $23\%$ just leads to an offset of the $L_{ii}^{(k)}/\lambda$ versus $Re_{\lambda}$ plots, analogous to what has been shown in \cite{VV2011}, figure 9.

Turning now to the results and starting with $L_{11}^{(1)}/\lambda$ and $L_{22}^{(1)}/\lambda$ versus $Re_{\lambda}$ for the centreline  (figure \ref{fig:Llambda}a) it is clear that both ratios are approximately constant throughout the assessed region of the decay ($1<x/x_{\mathrm{peak}}<4$).
This is the behaviour recently discovered for $L_{11}^{(1)}/\lambda$ by \cite{SV2007}, \cite{MV2010}, \cite{VV2011,VV2012}, \cite{gomesfernandesetal12} and \cite{discettietal11}. 
This behaviour is now found to extend to $L_{22}^{(1)}/\lambda$ (at least for the near-field decay region of RG115) and has been associated with nonequilibrium by \cite{VV2012}. 
A remarkable new finding which is reported here for the first time (in the near-field decay region of RG115) is that this behaviour occurs along three different streamwise lines with the same numerical constant for $L_{22}^{(1)}/\lambda$ (the
centreline $(y=0, z=0)$ and the lines $(y=-M/4, z=0)$ and $(y=-M/2, z=0)$) but not for $L_{11}^{(1)}/\lambda$ (see Figure \ref{fig:Llambda}a). 
Along the streamwise line $(y=-M/2, z=0)$ which lies within the wake of the lower bar of RG115's central mesh, $L_{11}^{(1)}/\lambda \sim Re_{\lambda}$ corresponding to the classical equilibrium behaviour where $C_{\varepsilon}^{1(1)}$ is independent of $Re_{\lambda}$. 
It is interesting that the classical equilibrium behaviour for $L_{11}^{(1)}/\lambda$ and $C_{\varepsilon}^{1(1)}$ is associated with $L_{22}^{(1)}/\lambda \sim \mathrm{constant}$ and $C_{\varepsilon}^{2(1)} \sim 1/Re_{\lambda}$ in the near-field field decay region of RG115 turbulence. 
Clearly the large eddies become less anisotropic as one probes them by moving downstream along the $(y=-M/2, z=0)$ line because $L_{11}^{(1)}/L_{22}^{(1)}$ decreases proportionally to $Re_{\lambda}$ as $Re_{\lambda}$ decreases. 
We do not know of any other relation such as $L_{11}^{(1)}/L_{22}^{(1)} \sim Re_{\lambda}$ in the literature to describe the large-scale anisotropy's dependence on $Re_{\lambda}$. 
It will be worth revisiting canonical free shear flows such as wakes and jets in future studies because, to our knowledge, only measurements of $C_{\varepsilon}^{1(1)}$ have been reported in such flows in support of $C_{\varepsilon}^{1(1)} \sim \mathrm{constant}$ for high enough Reynolds numbers \cite[e.g. see][]{Sreeni95,Pearson,Burattini2005}.
It will be interesting to know whether $C_{\varepsilon}^{2(1)} \sim 1/Re_{\lambda}$ and $L_{11}^{(1)}/L_{22}^{(1)} \sim Re_{\lambda}$ also hold in such flows or whether these relations are only valid in grid-generated turbulence. 
Note that a mixed behaviour $L_{11}^{(1)}/\lambda \approx A + B Re_{\lambda}$ ($A$ and $B$ are dimensionless constants) is observed along the intermediate streamwise line $(y=-M/4, 0)$ and that the downstream distances of our measurements relative to the bar thickness range within $95 < x/t_{0} <305$ which would typically be considered the far wake

Considering now the integral length-scales based on the transverse separations (figure \ref{fig:Llambda}b), the results show that $L_{22}^{(2)}/\lambda \approx \mathrm{constant}$ for both the centreline and behind the bar. 
This observation behind the bar also leads to the observation that $L_{11}^{(1)}/L_{22}^{(2)}\sim Re_{\lambda}$. 
On the other hand, at the centreline $L_{11}^{(2)}/\lambda$ increases  as the flow decays, which would imply that $C_{\varepsilon}^{1(2)}$ grows with decreasing $Re_{\lambda}$.

Note that, using the definitions of the Kolmogorov microscale, $\eta$($\equiv (\nu^3/\varepsilon)^{1/4}$), and of the Taylor microscale, it follows directly that $\ell/\eta \propto \ell/\lambda\, Re_{\lambda}^{1/2}$. 
Therefore, for $\ell/\lambda$ increasing faster than $Re_{\lambda}^{-1/2}$ as $Re_{\lambda}$ decreases, $\ell/\eta$ increases during decay.
We checked that $L_{11}^{(2)}/\lambda$ and $L/\lambda$ increase faster than $Re_{\lambda}^{-1/2}$ which leads to the unusual situation where $L_{11}^{(2)}/\eta$ and $L/\eta$ increase during decay. 
We are unable, at this point, to give a definitive explanation for this behaviour, but as discussed in \S \ref{sec:Lu} it may be related to periodic shedding from the bars which is contaminating the correlation functions, in particular $B_{11}^{(2)}$.

Lastly, the data from one-component longitudinal traverses for the several grids (see table \ref{table:summary}) are presented in figure \ref{fig:LlambdaB}. 
Here we use the isotropic definitions of the Taylor microscale $\lambda^{\mathrm{iso}}$($\equiv (15 \nu \overline{u^2}/\varepsilon)^{1/2}$) and $Re_{\lambda}^{\mathrm{iso}}$($\equiv u'\,\lambda/\nu$).
The RG115 and FSG3'x3' data behind the bar clearly follow $L_{11}^{(1)}/\lambda \propto Re_{\lambda}$, whereas $L_{11}^{(1)}/\lambda \approx \mathrm{constant}$ is a better approximation for the RG115 at the centreline.
For the FSG3'x3'  centreline data, on the other hand, a mixed behaviour, $L_{11}^{(1)}/\lambda \sim \mathrm{A} + \mathrm{B}\,Re_{\lambda}$, seems to be a better approximation, similar to the behaviour  of the RG115 data at $y=M/4$ (i.e. half way between the centreline and the bar, see figure \ref{fig:Llambda}a). 
(We confirmed that this behaviour is not due to any misalignments of the probe relative to the centreline.)
This was hypothesised in \cite{VV2011} to be a consequence of the confining walls, but as can be seen from the comparison between the FSG3'x3' versus FSG18''x18'' data and RG115 versus RG230 data, confinement does not meaningfully change the slope of $L_{11}^{(1)}/\lambda$ versus $Re_{\lambda}$ (c.f. figure \ref{fig:LlambdaB}).
The cause for this mixed behaviour is unclear, but it is plausible that it may be a consequence of the turbulence generated by the additional fractal iterations.
Note that, the smaller fractal iterations have a smaller wake-interaction length-scale, $x_*$ (see \S \ref{sec:wake}).
Extrapolating the observation from \cite{VV2012} that the extent of the nonequilibrium region is around $ 2\,x_*$, leads to the possibility that the turbulence generated by the two smallest fractal iterations transition earlier to equilibrium at $x\approx 1.2 x_{\mathrm{peak}}$ and $x\approx 2.2 x_{\mathrm{peak}}$, respectively, leading to the mixed behaviour. 
This is, however, no more than a tentative conceptual explanation at the moment which will require further investigation.

The centreline data for the RG60 is also included for reference in figure \ref{fig:LlambdaB}.  
It straddles both nonequilibrium and equilibrium turbulence, as can be educed from the change in slope from horizontal (from $Re_{\lambda}\approx 180$ to $Re_{\lambda}\approx 120$) to $L_{11}^{(1)}/\lambda \propto Re_{\lambda}$. 

\section{Main Conclusions}

The region of decaying grid-generated turbulence, where nonequilibrium
dissipation behaviour has been previously reported,
i.e. $x_{\mathrm{peak}}<x \lesssim 5\,x_{\mathrm{peak}}$, has been
experimentally investigated in the lee of both low-blockage RGs and
FSGs.

The geometrical differences between the grids can strongly influence
the transverse profiles of turbulent transport and production as well
as their downstream evolution. There are stark differences between the
FSGs and the RGs in the transverse profiles of the turbulent transport
throughout the assessed region of the flow.
The chocking due to the developing boundary layers on the confining
walls leads to distortions in the mean velocity and turbulent
production transverse profiles on some of our grids though not all of
them. This is an effect that is negligible upstream but becomes
increasingly noticeable far downstream, particularly if the test
section is excessively long for its width. 
Our findings demonstrate that the new nonequilibrium dissipation law $C_{\varepsilon} \sim Re_{M}^{m}/Re_{\ell}^{n}$ is present in the flow
irrespective of all these effects on the various transverse profiles and that $m \approx n \approx 1$ is a good approximation, in fact a little better for the RGs than for the FSGs.

We have also shown by studying the RG115-generated
turbulence in more detail, that the well-known equilibrium dissipation
law and the new nonequilibrium dissipation law can coexist in
different regions of the same flow. Specifically, the one-dimensional
surrogate of the normalised energy dissipation follows the previously
reported nonequilibrium form at the centreline,
i.e. $L_{11}^{(1)}/\lambda\approx \mathrm{constant}$ (equivalent to
$C_{\varepsilon}^{1(1)}\sim Re_{\lambda}^{-1}$ from \eqref{eq:Ceps}) ,
but behind the bar it follows the classical equilibrium law
$L_{11}^{(1)}/\lambda\sim Re_{\lambda}$ (equivalent to
$C_{\varepsilon}^{1(1)}\approx \mathrm{constant}$).  However, it is
perhaps remarkable that when the transverse integral scale is used,
the nonequilibrium form $L_{22}^{(1)}/\lambda\sim \mathrm{constant}$
is recovered both at and off centreline. This finding is significant
because (i) it implies that $L_{11}^{(1)}/L_{22}^{(1)}\sim
Re_{\lambda}$ in the lee of a bar of the grid and (ii) it indicates
that different integral length-scales can lead to very different time
scales and to apparently contradictory behaviours of the corresponding
surrogates of the normalised energy dissipation rate.

Using two-point/two-component measurements to obtain the correlation
functions for transverse separations ($B_{11}^{(2)}$ and
$B_{22}^{(2)}$) and their corresponding integral length-scales
($L_{11}^{(2)}$ and $L_{22}^{(2)}$), it is found that
$L_{22}^{(2)}/\lambda \approx \mathrm{constant}$ both along the
centreline and behind the bar. The observation behind the bar leads to
the conclusion that $L_{11}^{(1)}/L_{22}^{(2)}$ is also
proportional to $Re_{\lambda}$ along a straight line normal to the
grid crossing a bar of the grid.\\

P.C.V would like to thank Jovan Nedic for the help preparing the 3'x3' wind tunnel and acknowledges the financial support from Funda\c{c}\~{a}o para a Ci\^{e}ncia e a Tecnologia (grant SFRH/BD/61223/2009, cofinanced by POPH/FSE).

\bibliographystyle{jfm}
\bibliography{mybib}

\end{document}